\documentclass[twocolumn,showpacs,preprintnumbers,amsmath,amssymb]{revtex4}
\usepackage{graphicx}

\font\elevenmib=cmmib10 scaled 1095
\font\tenmib=cmmib10
\font\eightmib=cmmib10 scaled 800
\font\sixmib=cmmib10 scaled 667
\skewchar\elevenmib='177
\newfam\mibfam
\def\mib{\fam\mibfam\tenmib}
\textfont\mibfam=\tenmib
\scriptfont\mibfam=\eightmib
\scriptscriptfont\mibfam=\sixmib

\begin{document}
\title{Supersymmetric Valence Bond Solid States}
\author{Daniel P. Arovas}
\address{Department of Physics, University of California at San Diego,
La Jolla, California 92093, USA}
\author{Kazuki Hasebe}
\address{Department of General Education, Takuma National College of Technology, Takuma-cho, Mitoyo-city, Kagawa 769-1192, Japan }
\author{Xiao-Liang Qi and Shou-Cheng Zhang}
\address{Department of Physics, Stanford University, Stanford, California 94305, USA}

\date{\today}
\begin{abstract}
In this work we investigate the supersymmetric version of the
valence bond solid (SVBS) state. In one dimension, the SVBS states
continuously interpolate between the valence bond states for integer
and half-integer spin chains, and they generally describe
superconducting valence bond liquid states. Spin and superconducting
correlation functions can be computed exactly for these states, and
their correlation lengths are equal at the supersymmetric point. In higher dimensions,
the wave function for the SVBS states can describe resonating valence
bond states. The SVBS states for the spin models are shown to
be precisely analogous to the bosonic Pfaffian states of the quantum
Hall effect. We also give microscopic Hamiltonians for which the
SVBS state is the exact ground state.

\end{abstract}
\maketitle
\vskip2pc
\narrowtext

\tableofcontents
\section{Introduction}\label{sectintro}
Quantum antiferromagnetism offers basic paradigms for different
phases of strongly interacting quantum systems
\cite{villain1980od,Shender1982JETP,henley1989odd}. In addition to a
rich array of classically ordered states, including multiple
sublattice N{\'e}el order and non-collinear states, there are
several different types of quantum disordered states: valence bond (VB) 
solids, valence bond liquids, dimer solids, {\it etc}.  By tuning
various couplings, one can pass through quantum phase transitions
which separate these states. A class of superconductors, including
the high-$T_c$ cuprates, is obtained by doping the Mott insulating
states with quantum antiferromagnetic order. In one theoretical
approach, superconductivity arises from doping the valence bond
liquid state \cite{anderson1987rvb}. In another theoretical
approach, the superconducting state is obtained from a symmetry
rotation of the quantum antiferromagnetic state
\cite{Zhang1997science}. In this work we construct supersymmetric
extension of the valence bond solid state. In particular, we show
that the superconducting valence bond liquid state can be naturally
obtained from the supersymmetric rotation of the valence bond solid
state. Our results give a mathematical precise validation of the
above-mentioned ideas.

We investigate extensions of the valence bond solid states defined
by Affleck, Kennedy, Lieb, and Tasaki (AKLT) \cite{affleck1987rrv,affleck1988vbg}.
On any lattice ${\cal L} $, one can define a one-parameter family of such
states, indexed by an integer $M$.  The general AKLT state is
written \cite{arovas1988ehm}
\begin{equation}
\big|{{\rm{\Phi}}({\cal L} ,M)}\big\rangle=\prod_{\langle ij\rangle}
\big(\epsilon^{\vphantom{\dagger}}_{\mu\nu}\,
b^{\dagger}_{i\mu} b^{\dagger}_{j\nu}\big)^{\!M}\big|{0}\big\rangle ,
\label{akltstateoperator}
\end{equation}
as a product over links $\langle ij\rangle$ of ${\cal L} $, where
${\mib S} ^{\vphantom{\dagger}}_i=\frac{1}{2}
b^{\dagger}_{i\mu}{\mib\sigma} ^{\vphantom{\dagger}}_{\mu\nu}
b^{\vphantom{\dagger}}_{i\nu}$ is the local quantum spin operator,
written in terms of Schwinger bosons, satisfying $\big[
b^{\vphantom{\dagger}}_{i\mu},b^{\dagger}_{j\nu}\big]=\delta^{\vphantom{\dagger}}_{ij}\,\delta^{\vphantom{\dagger}}_{\mu\nu}$.
The state $\big|{{\rm{\Phi}}({\cal L} ,M)}\big\rangle$ describes an
antiferromagnet where each site contains a single spin
$S={\frac{1}{2}} z M$ object, where $z$ is the lattice coordination
number. What is special about these states is that the total spin
$J$ on any link can only take values between $0$ and $J^*\equiv
(z-1)M$, and all total spin components $J=J^*+1,\ldots,2S$ along any
link are absent in the AKLT wave function, because the operator
$\phi^{\dagger}_{ij}=\epsilon_{\mu\nu}\,b^{\dagger}_{i\mu}
b^{\dagger}_{j\nu}$ transforms as an \textsf{SU}(2) singlet. Thus,
the AKLT states are annihilated by certain projection operators, 
\begin{equation}
{\rm P}^{\vphantom{\dagger}}_J(ij)\big|{{\rm{\Phi}}({\cal L} ,M)}\big\rangle=0
\end{equation}
for $J\in\big[J^*+1,2S\big]$, where $J^*=(z-1)M$.  This allows one to construct local Hamiltonians of the form
\begin{equation}
H=\sum_{\langle ij\rangle}\sum_{J^*+1}^{2S} V^{\vphantom{\dagger}}_J\,{\rm P}^{\vphantom{\dagger}}_J(ij)\
,\label{AKLTHamiltonian}
\end{equation}
with $V^{\vphantom{\dagger}}_J\ge 0$, which renders $\big|{{\rm{\Phi}}({\cal L} ,M)}\big\rangle$ an exact, zero-energy
ground state.  In this respect, the AKLT states are analogous
to the Laughlin wave functions in the fractional quantum Hall effect (QHE) \cite{laughlin1983aqh}, which are also rendered exact eigenstates of a corresponding
``truncated pseudopotential'' Hamiltonian \cite{haldane1983fqh,arovas1988ehm}.
The $\textsf{SU}$(2) AKLT model has been generalized by introducing $q$-deformed \textsf{SU}(2) group \cite{klumper1991eas,klumper1992gpg,totsuka1994hsb}, and higher symmetric groups, such as $\textsf{SU}(n)$ \cite{affleck1991qae,arovas2008sss,greiter2007ers,greiter2007vbs}, $\textsf{SP}(n)$ \cite{schuricht2008vbs}, and $\textsf{SO}(n)$ \cite{Tu08041685,Tu08061839}.

The states we shall discuss are supersymmetric generalizations of
the AKLT states, and are written as
\begin{equation}
\big|{{\rm{\Psi}}({\cal L} ,M,r)}\big\rangle=\prod_{\langle ij\rangle}
\big(\epsilon{^{\vphantom{\sum}}}
_{\mu\nu}\,b^{\dagger}_{i\mu} b^{\dagger}_{j\nu} + r f^{\dagger}_i
f^{\dagger}_j\big)^M\big|{0}\big\rangle . \label{SAKLT}
\end{equation}
Here, $f^{\dagger}_i$ creates a fermionic hole on site $i$, which
displaces one of the bosons. The local Hilbert space thus
accommodates two types of states: states with spin $S={\frac{1}{2}}
z M$ and states with spin $S-{\frac{1}{2}}$, and the operator
${{\raise.35ex\hbox{$\chi$}}}^{\dagger}_{ij}=\epsilon_{\mu\nu}\,b^{\dagger}_{i\mu}
b^{\dagger}_{j\nu}+  r f^{\dagger}_i f^{\dagger}_j$, which creates a
linear combination of spin singlets and hole pairs on the link
$\langle ij\rangle$, transforms as a singlet under the superalgebra
$\textsf{OSp}(1|2)$.  We call these states supersymmetric valence
bond solid (SVBS) states. Physically, the spin $S$ states can be
realized by $2S$ electrons coupled through Hund's rule coupling, and
the spin $S-\frac{1}{2}$ states are obtained by removing one electron from
the site. Thus the SVBS states describe a doped spin chain with
large on-site Hund's rule coupling.

The parameter $r$ interpolates between two limits.  At $r=0$,
there are no holes, and we recover the AKLT state, which is an
antiferromagnetic insulator.  For finite $r$, there is a nonzero
density of nearest-neighbor hole pairs and the system is a
superconductor. The average spin per site is somewhere between
$S-{\frac{1}{2}}$ and $S$. As $r\to\infty$, each site must contain a hole,
and the state is insulating once again.   For a one-dimensional (1D) 
chain, with $M=1$, there are only two possibilities: 
\begin{subequations}
\begin{align}
\big|{{\rm\Phi}^{\vphantom{\dagger}}_{{\scriptscriptstyle{\rm A}}}  }\big\rangle
&=\big|{ {{\raise 0.5ex \hbox to 7pt{\leaders\hrule\hfill}}
\kern-1pt\hbox to 6pt{$\bullet$\hfill}}
  {{\hskip 13pt}}
{\hbox to 6pt{$\bullet$\hfill}\kern-1.5pt
{\raise 0.5ex \hbox to 14pt{\leaders\hrule\hfill}}
\kern-1pt\hbox to 6pt{$\bullet$\hfill}}{{\hskip 13pt}}
{\hbox to 6pt{$\bullet$\hfill}\kern-1.5pt
{\raise 0.5ex \hbox to 14pt{\leaders\hrule\hfill}}
\kern-1pt\hbox to 6pt{$\bullet$\hfill}}{{\hskip 13pt}}
{\hbox to 6pt{$\bullet$\hfill}\kern-1.5pt
{\raise 0.5ex \hbox to 14pt{\leaders\hrule\hfill}}
\kern-1pt\hbox to 6pt{$\bullet$\hfill}}{{\hskip 13pt}}
 {\hbox to 6pt{$\bullet$\hfill}\kern-1.5pt
{\raise 0.5ex \hbox to 7pt{\leaders\hrule\hfill}}}
\def\blank{{\hskip 13pt}} \ }\big\rangle\\
\big|{{\rm\Phi}^{\vphantom{\dagger}}_{{\scriptscriptstyle{\rm B}}}}\big\rangle&=
\big|{\ \ {\hbox to 6pt{$\bullet$\hfill}\kern-1.5pt
{\raise 0.5ex \hbox to 14pt{\leaders\hrule\hfill}}
\kern-1pt\hbox to 6pt{$\bullet$\hfill}}{{\hskip 13pt}}
{\hbox to 6pt{$\bullet$\hfill}\kern-1.5pt
{\raise 0.5ex \hbox to 14pt{\leaders\hrule\hfill}}
\kern-1pt\hbox to 6pt{$\bullet$\hfill}}{{\hskip 13pt}}
{\hbox to 6pt{$\bullet$\hfill}\kern-1.5pt
{\raise 0.5ex \hbox to 14pt{\leaders\hrule\hfill}}
\kern-1pt\hbox to 6pt{$\bullet$\hfill}}{{\hskip 13pt}}
{\hbox to 6pt{$\bullet$\hfill}\kern-1.5pt
{\raise 0.5ex \hbox to 14pt{\leaders\hrule\hfill}}
\kern-1pt\hbox to 6pt{$\bullet$\hfill}}\ \ \ }\big\rangle ,
\end{align}\label{twodegMGstates}
\end{subequations}
corresponding to spin-Peierls order.  These are the two degenerate ground states
of the well-known Majumdar-Ghosh Hamiltonian \cite{majumdar1969nnn,majumdar1970amk,BroekPLA1980}.
In the thermodynamic limit, or on a ring with an even number of sites, the $r\to\infty$ SVBS state is the sum $\big|{{\rm{\Phi}}^{\vphantom{\dagger}}_{{\scriptscriptstyle{\rm A}}}}\big\rangle+
\big|{{\rm{\Phi}}^{\vphantom{\dagger}}_{{\scriptscriptstyle{\rm B}}}}\big\rangle$, which has crystal momentum $k=0$.

\begin{figure}[!t]
\centering
\includegraphics[width=7.5cm]{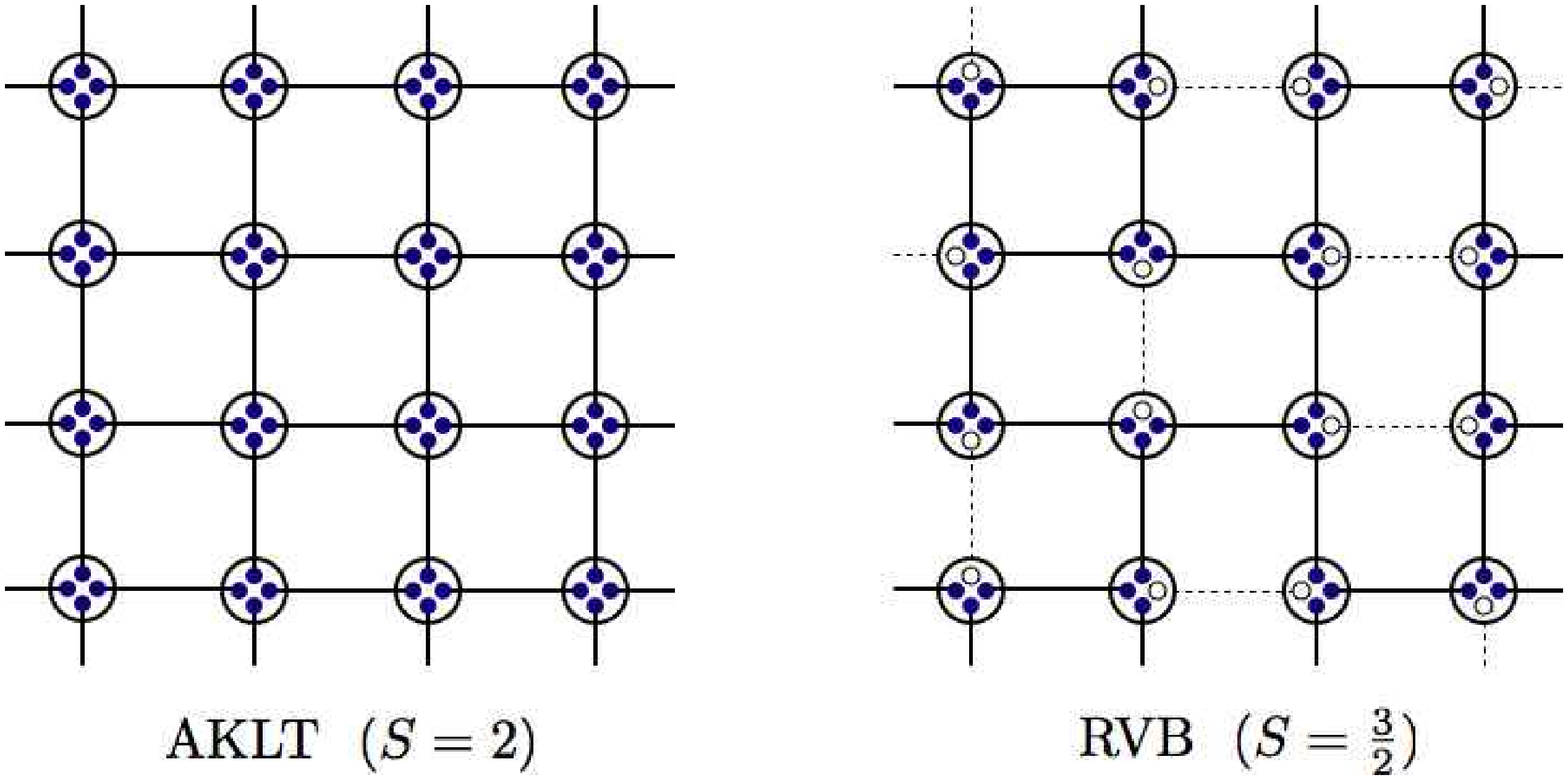}
\caption
{\label{AKLT_RVB} Examples of the square lattice supersymmetric valence bond solid state
with $M=1$ on the square lattice. Left panel: $r=0$, corresponding to the $S=2$ AKLT state.
Right panel: example configuration from the $r=\infty$ state which is a $S=\frac{3}{2}$ nearest-neighbor RVB state.}
\end{figure}

On the two-dimensional square lattice, once again the $r=0$ state is
the $S=2$ AKLT valence bond solid. For $r\to\infty$, though, rather
than there being only two configurations which contribute to the
SVBS wave function, the state is a linear combination of the
resonating valence bond (RVB) kind, but for $S=\frac{3}{2}$.  The
situation is depicted in Fig. \ref{AKLT_RVB}.   The configurations
which contribute to the SVBS state in this limit are given by dimer
coverings of the square lattice, where each dimer corresponds to a
hole-pair-creation operator $f^{\dagger}_i f^{\dagger}_j$.  The
quantum dimer model for $S=\frac{1}{2}$ was constructed by Rokhsar
and Kivelson \cite{rokhsar1988saq}. The partition function for the
classical dimer gas, with different fugacities for $x$-directed and
$y$-directed dimers, was worked out by Fisher in 1961 and
shown to take the form of a Pfaffian \cite{fisher1961smd}.  This
connection to the Pfaffian is present in our work as well, and
underlies recent work by one of us
\cite{hasebe2005PRL,hasebe2005qhl,hasebe2008ula} on supersymmetric
extensions of the quantum Hall problem, in which the Pfaffian QHE
state at $\nu={\frac{1}{2}}$ appears as a natural limit. The
following diagram sketches these basic connections:
\begin{center}
\begin{tabular}{ccccc}
AKLT ($S=1$ chain) && $\longrightarrow$ && Majumdar-Ghosh ($S={\frac{1}{2}}$)\\
&&&&\\
$\downarrow$ &&&& $\downarrow$ \\
&&&&\\
AKLT ($S=2$ square) && $\longrightarrow$ && RVB ($S=\frac{3}{2}$)\\
&&&&\\
$\downarrow$ &&&& $\downarrow$ \\
&&&&\\
Laughlin ($\nu={\frac{1}{2}}$ bosons)&& $\longrightarrow$ && Pfaffian ($\nu={\frac{1}{2}}$ fermions) \\
\end{tabular}
\end{center}

While the literature on hole motion in quantum antiferromagnets is
voluminous rather little has been done to date to explore
extensions of models of the AKLT type, {\it i.e.}, to find wave functions
at finite hole concentration which are exact eigenstates of local
projectors (so-called ``Klein models'' \cite{klein1982egs}).  Single hole
motion in the $S=1$ AKLT chain was discussed in Ref. \cite{zhang1989hms},
but a different constraint was used: $b^{\dagger}_\mu b^{\vphantom{\dagger}}_\mu +2 f^{\dagger}\!
f = 2$.
Experiments on hole doped AKLT spin chains have been
reported in Ref.\cite{xu2000hqs}, where a hole divides the
$S=1$ AKLT chain to two semi-indefinite segments with $S={1\over 2}$
spins at each edge. Interestingly, such property is shared in the
SVBS model developed in this paper.

The $t$-$J$ models with $\textsf{SU}(2|1)$ symmetry are known as the supersymmetric $t$-$J$ models. The models are exactly solvable in one dimension \cite{sutherland1975mmq,schlottmann1987inb,bares1990stm}, and their correlation functions are also derived in Ref.\cite{kawakami1990cfo}.
With $1/r^2$ long-range interaction, the supersymmetric $t$-$J$ models are still exactly solvable \cite{kuramoto1991ess}.
The models which we deal with possess $\textsf{OSp}(1|2)$ symmetry, and their exact ground states are constructed even in higher dimensions as the case of the original AKLT models.

The remainder of this paper is structured as follows.  In Sec.\ref{SUSY} we will briefly discuss the local Hilbert space and some
preliminary aspects of the $\textsf{OSp}(1|2)$ operator algebra, a
fuller treatment of which we consign to Appendix \ref{superalgebras}. In 
Sec.\ref{CHAINS} we will focus on SVBS states in one-dimensional
systems, {\it i.e.}, supersymmetric spin chains. Using the spin-hole
coherent states developed by Auerbach \cite{auerbach1994iea}, we
will compute various correlation functions in the SVBS chains.
Section \ref{QHE} discusses some connections with the quantum Hall
effect.  In Sec.\ref{trancpseudohamil}, we derive a Hamiltonian
with local interactions which renders our $M=1$ SVBS chain as an
exact nondegenerate ground state.

\section{Local Hilbert Space and $\textsf{OSp}(1|2)$}\label{SUSY}
In the Schwinger representation of $\textsf{SU}(2)$, a spin is represented by two bosons,
with the quantum spin operator given by ${\mib S} ={\frac{1}{2}} b^{\dagger}_\mu\,{\mib\sigma} ^{\vphantom{\dagger}}_{\mu\nu}\,b^{\vphantom{\dagger}}_\nu$.
The total boson occupancy is constrained,
\begin{equation}
a^{\dagger} a  + b^{\dagger} b =p\ ,
\end{equation}
where $p$ is an integer, and where we define $b^{\vphantom{\dagger}}_{\uparrow}
\equiv a$ and $b^{\vphantom{\dagger}}_{\downarrow}
\equiv b$.
The integer $p$ determines the representation of $\textsf{SU}(2)$; it corresponds to the number of columns in the corresponding Young diagram.  The spin magnitude is simply
$S={\frac{1}{2}}\, p$, and the dimension of the representation is $g=p+1$.

Now let us add in hole states.  The constraint equation becomes
\begin{equation}
a^{\dagger} a  + b^{\dagger} b  + f^{\dagger}\! f =p\ .
\label{gcon}
\end{equation}
There are now $g=2p+1$ possible states, corresponding to the two classes
\begin{align*}
S&={\frac{1}{2}} \,p\  &:&& a^{\dagger} a  + b^{\dagger} b &=p & \hbox{\rm and} &&
 f^{\dagger} f&=0\\
S&={\frac{1}{2}} \,(p-1)\  &:&&a^{\dagger} a  + b^{\dagger} b  &=p-1 & \hbox{\rm and}
&& f^{\dagger} f&=1\ .
\end{align*}
The simplest such case we shall deal with is $p=2$, for which $g=5$.  Explicitly, these states are given by
\begin{align*}
\big|{+1}\big\rangle&=\frac{1}{\sqrt{2}}\big(a^{\dagger}\big)^{\!2} \,{\big|{\textsf{V}}}\big\rangle \ ,&
\big|\!+\!\frac{1}{2}\big\rangle&=f^{\dagger} a^{\dagger} \,{\big|{\textsf{V}}}\big\rangle \ , \\
\big|{\,0\,}\big\rangle&=a^{\dagger} b^{\dagger}\,{\big|{\textsf{V}}}\big\rangle \ ,  & & \\
\big|{-1}\big\rangle&=\frac{1}{\sqrt{2}}\big(b^{\dagger}\big)^{\!2} \,{\big|{\textsf{V}}}\big\rangle \ , 
 & \big|-\!\frac{1}{2}\,\big\rangle&=f^{\dagger} b^{\dagger} \,{\big|{\textsf{V}}}\big\rangle
\ ,
\end{align*}
where ${\big|{\textsf{V}}}\big\rangle
$ is the vacuum for bosons and fermions: $a\,{\big|{\textsf{V}}\big\rangle}
=b\,{\big|{\textsf{V}}\big\rangle}
=f\,{\big|{\textsf{V}}\big\rangle}
=0$.

The $2p+1$ states obeying the constraint of Eq.(\ref{gcon}) may be grouped into
a multiplet of the superalgebra $\textsf{OSp}(1|2)$.  This group has five generators, three
of which are the familiar $\textsf{SU}(2)$ spin operators: $L^{\vphantom{\dagger}}_a={\frac{1}{2}}\,b^{\dagger}_\mu\,\sigma^a_{\mu\nu}\,
b^{\vphantom{\dagger}}_\nu$, with $i=1,2,3$.  The remaining two generators are non-Hermitian, and may be taken to be
\begin{align}
K^{\vphantom{\dagger}}_1&={\frac{1}{2}}\big(x^{-1}f^{\vphantom{\dagger}} a^{\dagger} + x f^{\dagger} b\big)\ , \nonumber\\
K^{\vphantom{\dagger}}_2&={\frac{1}{2}}\big(x^{-1}f^{\vphantom{\dagger}} b^{\dagger} -x f^{\dagger} a\big)\ ,
\end{align}
where $x$ is an arbitrary complex number.   The relations among the generators are
\begin{align}
\big[L^{\vphantom{\dagger}}_a\, , \, L^{\vphantom{\dagger}}_b\big]&=i\epsilon^{\vphantom{\dagger}}_{abc}\,L^{\vphantom{\dagger}}_c \ ,\nonumber\\
\big[L^{\vphantom{\dagger}}_a\, , \, K^{\vphantom{\dagger}}_\mu\big]&={\frac{1}{2}}\sigma^a_{\nu\mu}\,K^{\vphantom{\dagger}}_\nu \ ,\nonumber\\
\big\{K^{\vphantom{\dagger}}_\mu \, , \, K^{\vphantom{\dagger}}_\nu\big\}&={\frac{1}{2}}\big(i\sigma^y \sigma^a\big)^{\vphantom{\dagger}}_{\mu\nu}\,L^{\vphantom{\dagger}}_a\ .
\end{align}
The algebra of the generators is independent of the parameter $x$.  Note that $K_1^2={\frac{1}{4}} a^{\dagger} b={\frac{1}{4}} L^{\vphantom{\dagger}}_+$,
so $K^{\vphantom{\dagger}}_1$ is like the ``square root'' of the angular momentum raising operator $L^{\vphantom{\dagger}}_+=L^{\vphantom{\dagger}}_1+iL^{\vphantom{\dagger}}_2$.
Since $\big[L^{\vphantom{\dagger}}_3\, , \, K^{\vphantom{\dagger}}_1\big]={\frac{1}{2}} K^{\vphantom{\dagger}}_1$, we have that $K^{\vphantom{\dagger}}_1$ raises $L^{\vphantom{\dagger}}_3$ by half.
Similarly, $K^{\vphantom{\dagger}}_2$ lowers $L^{\vphantom{\dagger}}_3$ by half, and functions as the square root of the angular-momentum
lowering operator $L^{\vphantom{\dagger}}_-$.

The Casimir operator is given by
\begin{equation}
{\cal C} ={\mib L} ^2 + \epsilon_{\mu\nu}\,K^{\vphantom{\dagger}}_\mu K^{\vphantom{\dagger}}_\nu\ .
\end{equation}
Acting on the single-site states defined above, ${\cal C} $ takes the
value ${\frac{1}{4}} p(p+1)$. Generally, one has ${\cal C} =L(L+{\frac{1}{2}})$, where
$L$, which is either integer or half odd integer, is the maximum
eigenvalue of $L^{\vphantom{\dagger}}_3$, for a given value of ${\cal C} $.  We call this quantity $L$ the angular momentum.  The
dimension of the representation with angular momentum $L$ is
$g=4L+1$. The addition of angular momenta within
$\textsf{OSp}(1|2)$ is similar to the $\textsf{SU}(2)$ case,
except the spacing between consecutive $L$ values is ${\frac{1}{2}}$
rather than $1$,
\begin{equation}
L \otimes L' = |L-L'|  \oplus \left(|L - L'|+ {\frac{1}{2}}
\right)\oplus \cdots\oplus (L + L') \ .
\end{equation}
For example, if the local representation
of $\textsf{OSp}(1|2)$ on a single site is $L=1$, then on any link $(ij)$, one can have
\begin{equation}
1\otimes 1 = 0\oplus{\frac{1}{2}}\oplus 1\oplus\frac{3}{2}\oplus 2\ ,
\label{osp1times1}
\end{equation}
where the dimensions of the five irreducible representations in the product
are $1$, $3$, $5$, $7$, and $9$.
The Casimir operator for the two-site system is
\begin{align}
{\cal C} (ij)&=2{\mib L} (i) \cdot {\mib L} (j) + {\frac{1}{2}} F^{\dagger}_{ij} \, f^{\dagger}_j f^{\vphantom{\dagger}}_i +
{\frac{1}{2}} F^{\vphantom{\dagger}}_{ij} \,f^{\dagger}_i f^{\vphantom{\dagger}}_j\label{susyspinspinint}\\
&\qquad +{\frac{1}{2}} x^{-2} A^{\dagger}_{ij} \,f^{\vphantom{\dagger}}_i f^{\vphantom{\dagger}}_j +{\frac{1}{2}} x^2 A^{\vphantom{\dagger}}_{ij} \,f^{\dagger}_i f^{\dagger}_j
+{\cal C} (i)+{\cal C} (j)\ , \nonumber
\end{align}
where
\begin{subequations}
\begin{align}
A^{\vphantom{\dagger}}_{ij}&=a^{\vphantom{\dagger}}_i b^{\vphantom{\dagger}}_j - b^{\vphantom{\dagger}}_i a^{\vphantom{\dagger}}_j \\
F^{\vphantom{\dagger}}_{ij}&=a^{\dagger}_i a^{\vphantom{\dagger}}_j + b^{\dagger}_i b^{\vphantom{\dagger}}_j\ .
\end{align}
\end{subequations}
Using the Casimir operator, we can construct projection operators onto representations of a desired value of $L$.
This can be used to construct a Hamiltonian along the lines of AKLT; this program is carried out in Sec.\ref{trancpseudohamilsvbs}.  The link operator
\begin{equation}
{{\raise.35ex\hbox{$\chi$}}}^{\dagger}_{ij}=a^{\dagger}_i b^{\dagger}_j - b^{\dagger}_i a^{\dagger}_j +r f^{\dagger}_i f^{\dagger}_j
\end{equation}
transforms as an $\textsf{OSp}(1|2)$ singlet whenever $x^2=-r$.  That is to say
$\big[ L^{\vphantom{\dagger}}_a\, , \, {{\raise.35ex\hbox{$\chi$}}}^{\dagger}_{ij}\big]=\big[ K^{\vphantom{\dagger}}_\mu\, , \, {{\raise.35ex\hbox{$\chi$}}}^{\dagger}_{ij}\big]=0$
whenever $x=\pm i r$, where $L^{\vphantom{\dagger}}_a=\sum_i L^{\vphantom{\dagger}}_a(i)$ and
$K^{\vphantom{\dagger}}_\mu=\sum_i K^{\vphantom{\dagger}}_\mu(i)$ are global generators.   Thus, if each site is in the $L$
representation, there are $2L$ quanta per site, with $p=2L$ in Eq.(\ref{gcon}).
There are thus $4L$ quanta on each link.  In the general SUSY AKLT state of Eq.(\ref{SAKLT}),
$2M$ of these quanta are passivated in singlet bonds.  Thus, the maximum value of $L^{\vphantom{\dagger}}_{ij}$
for the link is $J^{\vphantom{\dagger}}_{{\scriptscriptstyle{\rm max}}}=2L-M$, where $L={\frac{1}{2}} z M$ relates the value of $L$, the lattice
coordination number $z$, and the integer parameter $M$.  For the $L=1$ SVBS chain, for example,
the wave function is annihilated by projectors onto either of the $L_{i,i+1}=\frac{3}{2}$ or $L_{i,i+1}=2$ sectors,
and the only remaining possibilities are $L^{\vphantom{\dagger}}_{i,i+1}=0$, ${\frac{1}{2}}$, and $1$.

However, an inconvenient problem remains.  Because the generators $K^{\vphantom{\dagger}}_\mu$ are
non-Hermitian, this is also the case for the projection operators, and, thus, the Hamiltonian as well.
Then, we use a ``trick'' to make a Hermitian Hamiltonian from non-Hermitian projection operators as performed in Sec.\ref{trancpseudohamil}.
In Sec.\ref{MODEL}, we also exhibit a properly Hermitian Hamiltonian which has the
$L=1$ SVBS chain $\it{with}$  fixed total fermion number an exact nondegenerate ground state.  Before doing so, we will derive the
properties of the SVBS chains themselves.

\section{SVBS Spin Chains}\label{CHAINS}
The general SUSY AKLT chain wave function is written as the pair product,
\begin{equation}
\big|{\rm{\Psi}}\big\rangle
=\prod_i\big(a^{\dagger}_i\, b^{\dagger}_{i+1} - b^{\dagger}_i
a^{\dagger}_{i+1}
+ r\,f^{\dagger}_i \,f^{\dagger}_{i+1}\big)^M\,{\big|{\textsf{V}}\big\rangle}
\ .\label{generalakltchainfunc}
\end{equation}
This describes a chain in which each site is in the $L=M$ representation of $\textsf{OSp}(1|2)$.
The wave function is annihilated by projectors ${\rm P}^{\vphantom{\dagger}}_{\!\!J}(i,i+1)$ which project onto the total
link angular momentum $J$, for $M<J\le 2M$.

We are interested in computing correlation functions in these states.  The correlation functions
we will compute are:
\begin{subequations}
\begin{align}
C^{\vphantom{\dagger}}_{{\scriptscriptstyle{\rm spin}}}(n)&=\big\langle \,{\mib L} (j)\cdot{\mib L} (j+n) \, \big\rangle \ , \\
C^{\vphantom{\dagger}}_{{\scriptscriptstyle{\rm SS}}}(n)&=\big\langle \big(a^{\vphantom{\dagger}}_j\, b^{\vphantom{\dagger}}_{j+n}-b^{\vphantom{\dagger}}_j\, a^{\vphantom{\dagger}}_{j+n}\big) f^{\dagger}_j\, f^{\dagger}_{j+n}\,\big\rangle\ ,
\end{align}
\end{subequations}
corresponding to the spin-spin correlation function and a ``singlet
superconductivity'' order-parameter correlation function. Since our
state does not conserve particle number, the superconducting order
parameter can be non-vanishing. Here $\langle {\cal O}  \rangle =
\big\langle{{\rm{\Psi}}}\big|{{\cal O} }\big|{{\rm{\Psi}}}\big\rangle\big/
\big\langle{{\rm{\Psi}}}\,\big|\,{{\rm{\Psi}}}\big\rangle$ is the
normalized expectation value.   A corresponding ``triplet 
superconductivity'' correlator,
\begin{equation}
C^a_{{\scriptscriptstyle{\rm TS}}}(n)=\big\langle \begin{pmatrix} a^{\vphantom{\dagger}}_j \, a{^{\vphantom{\sum}}}
_{j+n}  \\
{1\over\sqrt{2}}\big(a{^{\vphantom{\sum}}}
_j\,b{^{\vphantom{\sum}}}
_{j+n} + b{^{\vphantom{\sum}}}
_j\,a{^{\vphantom{\sum}}}
_{j+n}\big) \\ b^{\vphantom{\dagger}}_j\,b{^{\vphantom{\sum}}}
_{j+n}
\end{pmatrix}\,f^{\dagger}_j\,f^{\dagger}_{j+n}\,\big\rangle\ ,
\end{equation}
may also be defined.  However, due to the singlet property of the
SVBS states, we have that $C^a_{{\scriptscriptstyle{\rm TS}}}(l)=0$.  We shall compute
these correlations on finite chains, which have ends, and examine
the thermodynamic limit.  There are some specific properties of
edge states in these models, in direct correspondence to what is
known from AKLT chains \cite{kennedy1990edo,hagiwara1990osd,glarum1991ofs}. For example, the edges
of the $L=1$ SVBS chain are local $L={\frac{1}{2}}$ degrees of freedom,
which means that the ground state of a long but finite $L=1$ SVBS
chain will exhibit a ninefold quasi-degeneracy, with the actual
levels arranged into singlet, triplet, and quintuplet states,
according to ${\frac{1}{2}}\otimes{\frac{1}{2}} = 0\oplus {\frac{1}{2}} \oplus 1$.

Note that the operators whose correlation functions are computed must commute with the
local constraint $n^{\vphantom{\dagger}}_a + n^{\vphantom{\dagger}}_b + n^{\vphantom{\dagger}}_f=p$.  Expressions such as $\langle f^{\dagger}_j \,f^{\dagger}_{j+n}\rangle$
and $\langle a^{\vphantom{\dagger}}_j\, b^{\vphantom{\dagger}}_{j+n}\rangle$ vanish identically.

\subsection{Spin-hole coherent states}
The application of spin-coherent states in elucidating the
properties of the AKLT VBS states was discussed in Ref.
\cite{arovas1988ehm}.  Here we utilize a generalization of the familiar
$\textsf{SU}(2)$ spin-coherent states, known as spin-hole coherent
states \cite{auerbach1994iea}.  Define the state
\begin{align}
\big|{{\hat{\mib n}},\theta;p}\big\rangle&\equiv {1\over\sqrt{p\,!}}\big(u a^{\dagger} + v b^{\dagger} -\theta f^{\dagger}\big)^p {\big|{\textsf{V}}\big\rangle}
\nonumber\\
&=\big|{\hat{\mib n}}\big\rangle^{\vphantom{\dagger}}_{p}\otimes
\big|{0}\big\rangle -\sqrt{p} \, \theta\big|{{\hat{\mib n}}}\big\rangle^{\vphantom{\dagger}}_{p-1}\otimes \big|{1}\big\rangle \ .
\end{align}
Here, $\big|{\hat{\mib n}}^{\vphantom{\dagger}}_p\big\rangle$ is an $\textsf{SU}(2)$ spin-coherent state with $S={\frac{1}{2}} p$, and $\theta$
is a Grassmann variable which anticommutes with $f$ and $f^{\dagger}$.  The resolution of the identity
may be written as
\begin{equation}
\int\!\!{d{\hat{\mib n}}\over 4\pi}\!\!\int\!\!d{{\bar\theta}}\!\!\int\!\!d\theta\,e^{(p+1)\theta{{\bar\theta}}}\,\big|{{\hat{\mib n}},\theta;p}\,\big\rangle\,\big\langle{{\hat{\mib n}},\theta;p}|=
{\rm P}{^{\vphantom{\sum}}}_{L={p\over 2}}\ ,
\label{resid}
\end{equation}
where ${\rm P}{^{\vphantom{\sum}}}
_L$ is the projector onto the angular momentum $L$ representation of $\textsf{OSp}(1|2)$.

Next, consider a general state in the angular momentum $L$ representation, written as
\begin{align}
\big|{\psi}\big\rangle&={1\over\sqrt{p\,!}}\psi(a^{\dagger}\, , \, b^{\dagger} \, , \, f^{\dagger})\,{\big|{\textsf{V}}\big\rangle}
\\
&={1\over\sqrt{p\,!}}\Big[\psi^{\vphantom{\dagger}}_p(a^{\dagger}\, , \, b^{\dagger}) + \psi^{\vphantom{\dagger}}_{p-1}(a^{\dagger}\, , \, b^{\dagger})\,f^{\dagger}\Big]{\big|{\textsf{V}}\big\rangle}
\ ,\nonumber
\end{align}
where $\psi^{\vphantom{\dagger}}_p(a^{\dagger}\, , \, b^{\dagger})$ is homogeneous of degree $p$ in $a^{\dagger}$ and $b^{\dagger}$.  We then have
\begin{equation}
\big\langle{{\hat{\mib n}},\theta;p}\big|{\psi}\big\rangle=\psi({{\bar u}},{{\bar v}},{{\bar\theta}})\ .
\end{equation}
That is, we simply replace $a^{\dagger}\to{{\bar u}}$, $b^{\dagger}\to{{\bar v}}$, and $f^{\dagger}\to{{\bar\theta}}$ in the function $\psi$.

\subsection{Matrix elements}
Now consider the following spin operators:
\begin{subequations}
\begin{align}
{\hat T}^0_k&=\sum_{m=0}^k\sum_{n=0}^k T^0_{kmn}\,
 {a^{\vphantom{\dagger}}}^m\, {b^{\vphantom{\dagger}}}^{k-m}\, {{a^{\dagger}}}^n\, {{b^{\dagger}}}^{k-n} \ ,\\
 {\hat T}^+_k&=\sum_{m=0}^k\sum_{n=0}^{k+1} T^+_{kmn}\,
 {a^{\vphantom{\dagger}}}^m\, {b^{\vphantom{\dagger}}}^{k-m}\, {{a^{\dagger}}}^n\, {{b^{\dagger}}}^{k+1-n}\ ,\\
  {\hat T}^-_k&=\sum_{m=0}^{k+1}\sum_{n=0}^k T^-_{kmn}\,
 {a^{\vphantom{\dagger}}}^m\, {b^{\vphantom{\dagger}}}^{k+1-m}\, {{a^{\dagger}}}^n\, {{b^{\dagger}}}^{k-n}\ .
\end{align}
\end{subequations}
Note that ${\hat T}^\pm_k$ raise ($+$) or lower ($-$) the angular momentum by ${\rm{\Delta}} L={\frac{1}{2}}$,
while ${\hat T}^0_k$ preserves total spin.  Our goal is to compute the matrix element
\begin{align}
\big\langle{\psi}\big|{{\hat{\cal T} }_k}\big|{\phi}\big\rangle&={1\over p\,!}\,{\big\langle{\textsf{V}}}\big|
  {\bar\psi}(a,b,f)\,\Big[
{\hat T}^0_k\, + {\hat T}^+_k\, f \\
&\qquad\quad+ {\hat T}^-_k\, f^{\dagger}  + {\hat T}^{0'}_k\, ff^{\dagger}\Big]\,\phi(a^{\dagger},b^{\dagger},f^{\dagger}) {\big|{\textsf{V}}\big\rangle}%
\nonumber
\end{align}
and to represent it as an integral over spin-hole coherent states.   We find
\begin{align}
\big\langle{\psi}\big|{{\hat{\cal T} }_k}\big|{\phi}\big\rangle&=\int\!{d{\hat{\mib n}}\over 4\pi}\!\!
\int\!\!d{{\bar\theta}}\!\!\int\!\!d\theta\,e^{(p+1)\theta{{\bar\theta}}}\nonumber\\
&\qquad\times{\bar\psi}(u,v,\theta)\,
{\cal T} ^{\vphantom{\dagger}}_k({\hat{\mib n}},\theta,{{\bar\theta}})\,\phi({{\bar u}},{{\bar v}},{{\bar\theta}})\ ,
\end{align}
where
\begin{align}
{\cal T} ^{\vphantom{\dagger}}_k({\hat{\mib n}},\theta,{{\bar\theta}})&={(p\!+\!k\!+\!1)!\over p\,!}\,\Bigg[{T^0_k({\hat{\mib n}})
\over p+k+1} + T^+_k({\hat{\mib n}})\,\theta \label{repa}\\
&\qquad+ T^-_k({\hat{\mib n}})\,{{\bar\theta}} + T^{0'}_k({\hat{\mib n}})\,\theta{{\bar\theta}}
\,\Bigg]\,e^{k\theta{{\bar\theta}}}\nonumber
\end{align}
replaces
\begin{equation}
{\hat{\cal T} }^{\vphantom{\dagger}}_k={\hat T}^0_k\, + {\hat T}^+_k\, f + {\hat T}^-
_k\, f^{\dagger}  +
{\hat T}^{0'}_k\, ff^{\dagger}\ .
\label{repb}
\end{equation}

\subsection{Correlation functions}
With the spin-hole coherent state formalism developed, we are now in position to calculate the
correlation functions in the general SVBS chain state.  The first step is to compute the wave function
normalization, which we call ${\cal D}$ (for ``denominator'').  Using the resolution of unity for the spin-hole
coherent states, we have
\begin{align}
{\cal D}&=\big\langle{{\rm{\Psi}}}\big|{{\rm{\Psi}}}\big\rangle\nonumber\\
&=\int\!\!d\mu\,\prod_{n=0}^N \big|u^{\vphantom{\dagger}}_n\,v^{\vphantom{\dagger}}_{n+1} - v^{\vphantom{\dagger}}_n\,u^{\vphantom{\dagger}}_{n+1} + r\,\theta^{\vphantom{\dagger}}_{n+1}\,\theta^{\vphantom{\dagger}}_n\big|^2\ ,
\end{align}
where the measure is
\begin{align}
d\mu&=\prod_{j=0}^{N+1} \Bigg[{d{\hat{\mib n}}^{\vphantom{\dagger}}_j\over 4\pi}\,d{{\bar\theta}}^{\vphantom{\dagger}}_j\,d\theta^{\vphantom{\dagger}}_j\Bigg]\,
e^{(M+1)(\theta^{\vphantom{\dagger}}_0{{\bar\theta}}^{\vphantom{\dagger}}_0+\theta^{\vphantom{\dagger}}_{N+1}{{\bar\theta}}^{\vphantom{\dagger}}_{N+1})}\nonumber\\
&\qquad\qquad\qquad\times e^{(2M+1)(\theta^{\vphantom{\dagger}}_1{{\bar\theta}}^{\vphantom{\dagger}}_1+\ldots\theta^{\vphantom{\dagger}}_N{{\bar\theta}}^{\vphantom{\dagger}}_N)}\ .
\end{align}
Note that the site $j=0$ and $j=N+1$, which are at the ends of the chain and have only one neighbor,
are in the $L={\frac{1}{2}} M$ representation of $\textsf{OSp}(1|2)$ while the bulk sites are in the $L=M$ representation.
We now expand 
\begin{align}
&\Big|u^{\vphantom{\dagger}}_n\,v^{\vphantom{\dagger}}_{n+1} - v^{\vphantom{\dagger}}_n\,u^{\vphantom{\dagger}}_{n+1} + r\,\theta^{\vphantom{\dagger}}_{n+1}\,\theta^{\vphantom{\dagger}}_n\Big|^2=
\bigg({1-{\hat{\mib n}}^{\vphantom{\dagger}}_n\!\cdot{\hat{\mib n}}^{\vphantom{\dagger}}_{n+1}\over 2}\bigg)^{\!M} \nonumber\\
&\qquad+M r\bigg({1-{\hat{\mib n}}^{\vphantom{\dagger}}_n\!\cdot{\hat{\mib n}}^{\vphantom{\dagger}}_{n+1}\over 2}\bigg)^{\!M-1}
({{\bar u}}^{\vphantom{\dagger}}_n\,{{\bar v}}^{\vphantom{\dagger}}_{n+1} - {{\bar v}}^{\vphantom{\dagger}}_n\,{{\bar u}}^{\vphantom{\dagger}}_{n+1})\,\theta^{\vphantom{\dagger}}_{n+1}\,\theta^{\vphantom{\dagger}}_n\nonumber\\
&\qquad+M {{\bar r}}\bigg({1-{\hat{\mib n}}^{\vphantom{\dagger}}_n\!\cdot{\hat{\mib n}}^{\vphantom{\dagger}}_{n+1}\over 2}\bigg)^{\!M-1}
(u^{\vphantom{\dagger}}_n\,v^{\vphantom{\dagger}}_{n+1} - v^{\vphantom{\dagger}}_n\,u^{\vphantom{\dagger}}_{n+1})\,{{\bar\theta}}^{\vphantom{\dagger}}_n\,{{\bar\theta}}^{\vphantom{\dagger}}_{n+1}\nonumber\\
&\qquad+M^2 |r|^2 \bigg({1-{\hat{\mib n}}^{\vphantom{\dagger}}_n\!\cdot{\hat{\mib n}}^{\vphantom{\dagger}}_{n+1}\over 2}\bigg)^{\!M-1} \theta^{\vphantom{\dagger}}_n\,{{\bar\theta}}^{\vphantom{\dagger}}_n
\,\theta^{\vphantom{\dagger}}_{n+1}\,{{\bar\theta}}^{\vphantom{\dagger}}_{n+1}\ .\label{expand}
\end{align}
Using
\begin{equation}
\int\!{d{\hat{\mib n}}\over 4\pi}\,\bigg({1-{\hat{\mib n}}\cdot{\hat{\mib n}}'\over 2}\bigg)^{\!M}={1\over M+1}\ ,
\end{equation}
we can now integrate out site $j=0$.  The new integrand is then the truncated product wave function,
starting with site $j=1$, multiplied by the quantity $\alpha^{\vphantom{\dagger}}_1 + \beta^{\vphantom{\dagger}}_1\theta^{\vphantom{\dagger}}_1{{\bar\theta}}^{\vphantom{\dagger}}_1$,
where $\alpha^{\vphantom{\dagger}}_1=1$ and $\beta^{\vphantom{\dagger}}_1=M |r|^2$.  The form of this expression self-replicates.
That is, after integrating out sites $j=0$ through $j=n-1$ in succession, we are left with the
expression $\alpha^{\vphantom{\dagger}}_n + \beta^{\vphantom{\dagger}}_n\theta^{\vphantom{\dagger}}_n{{\bar\theta}}^{\vphantom{\dagger}}_n$.  We can now integrate out site $n$
to obtain the replication formula, 
\begin{align}
&\alpha^{\vphantom{\dagger}}_{n+1} + \beta^{\vphantom{\dagger}}_{n+1}\,\theta^{\vphantom{\dagger}}_{n+1}{{\bar\theta}}^{\vphantom{\dagger}}_{n+1}\equiv
\int\!\!{d{\hat{\mib n}}^{\vphantom{\dagger}}_n\over 4\pi}\!\!\int\!\!d{{\bar\theta}}^{\vphantom{\dagger}}_n\!\!\int\!\!d\theta^{\vphantom{\dagger}}_n\,
e^{(2M+1)\theta^{\vphantom{\dagger}}_n{{\bar\theta}}^{\vphantom{\dagger}}_n}\nonumber\\
&\qquad\times(\alpha^{\vphantom{\dagger}}_n + \beta^{\vphantom{\dagger}}_n\,\theta^{\vphantom{\dagger}}_n{{\bar\theta}}^{\vphantom{\dagger}}_n)
\times \Bigg[\bigg({1-{\hat{\mib n}}^{\vphantom{\dagger}}_n\cdot{\hat{\mib n}}^{\vphantom{\dagger}}_{n+1}\over 2}\bigg)^{\!M} \nonumber\\
&\qquad+M^2 |r|^2\,\bigg({1-{\hat{\mib n}}^{\vphantom{\dagger}}_n\cdot{\hat{\mib n}}^{\vphantom{\dagger}}_{n+1}\over 2}\bigg)^{\!M-1}\,
\theta^{\vphantom{\dagger}}_n{{\bar\theta}}^{\vphantom{\dagger}}_n\,\theta^{\vphantom{\dagger}}_{n+1}{{\bar\theta}}^{\vphantom{\dagger}}_{n+1}\Bigg]\nonumber\\
&\quad=\bigg({2M+1\over M+1}\,\alpha^{\vphantom{\dagger}}_n + {1\over M+1}\,\beta^{\vphantom{\dagger}}_n\bigg)
+ M|r|^2\,\alpha^{\vphantom{\dagger}}_n\,\theta^{\vphantom{\dagger}}_{n+1}{{\bar\theta}}^{\vphantom{\dagger}}_{n+1}\ .\label{bige}
\end{align}
Note that in propagating the expression $(\alpha^{\vphantom{\dagger}}_n+\beta^{\vphantom{\dagger}}_n
\theta^{\vphantom{\dagger}}_n{{\bar\theta}}^{\vphantom{\dagger}}_n)$, we may drop the last two terms on the right-hand-side (RHS) of Eq.(\ref{expand}).
We now have
\begin{equation}
\begin{pmatrix}\alpha^{\vphantom{\dagger}}_{n+1} \\ \\ \beta^{\vphantom{\dagger}}_{n+1}\end{pmatrix}=
\stackrel{{\cal D} }{\overbrace{\begin{pmatrix} \frac{2M+1}{M+1} && \frac{1}{M+1} \\ && \\
M|r|^2 && 0 \end{pmatrix}}}\begin{pmatrix}\alpha^{\vphantom{\dagger}}_n \\ \\ \beta^{\vphantom{\dagger}}_n\end{pmatrix}\ .
\label{deqn}
\end{equation}
When we get to the last site, we have the final result
\begin{align}
{\cal D}&=\int\!\!{d{\hat{\mib n}}^{\vphantom{\dagger}}_{N+1}\over 4\pi}\!\!\int\!\!d{{\bar\theta}}^{\vphantom{\dagger}}_{N+1}\!\!\int\!\!
d\theta^{\vphantom{\dagger}}_{N+1}\,e^{(M+1)\theta^{\vphantom{\dagger}}_{N+1}{{\bar\theta}}^{\vphantom{\dagger}}_{N+1}}\nonumber\\
&\qquad\qquad\times\big(\alpha^{\vphantom{\dagger}}_{N+1}+ \beta^{\vphantom{\dagger}}_{N+1}\,\theta^{\vphantom{\dagger}}_{N+1}{{\bar\theta}}^{\vphantom{\dagger}}_{N+1}\big)\nonumber\\
&=(M\!+\!1)\,\alpha^{\vphantom{\dagger}}_{N+1}+ \beta^{\vphantom{\dagger}}_{N+1}\ .
\end{align}
Thus,
\begin{equation}
{\cal D}=\begin{pmatrix} M\!+\!1 &  1\end{pmatrix}
{\cal D} ^N \begin{pmatrix} 1 \\ M|r|^2\end{pmatrix}\ .
\end{equation}
Now we need to compute the numerator for the correlation function of interest.

\subsubsection{Singlet superconductivity correlations}
We define the singlet off-diagonal correlation function
\begin{equation}
C^{\vphantom{\dagger}}_{{\scriptscriptstyle{\rm SS}}}(n)=\langle \,\frac{1}{\sqrt{2}}
(a^{\vphantom{\dagger}}_k\,b^{\vphantom{\dagger}}_{k+n} - b^{\vphantom{\dagger}}_k\,a^{\vphantom{\dagger}}_{k+n})\,f^{\dagger}_k\,f^{\dagger}_{k+n}\,\rangle\ ,
\end{equation}
which is independent of $k$ in the limit of a long chain
($N\to\infty$). The operator above, in the language of the
operators ${\hat T}^\sigma_k$ studied earlier, is of the form
${\hat T}^-_{k=0}$ on sites $k$ and $k+n$. Invoking Eq.
(\ref{repa}), we have
\begin{align}
&(a^{\vphantom{\dagger}}_k\,b^{\vphantom{\dagger}}_{k+n} - b^{\vphantom{\dagger}}_k\,a^{\vphantom{\dagger}}_{k+n})\,f^{\dagger}_k\,f^{\dagger}_{k+n}\to\\
&\qquad\qquad(2M+1)^2\,(u^{\vphantom{\dagger}}_k\,v^{\vphantom{\dagger}}_{k+n}-v^{\vphantom{\dagger}}_k\,u^{\vphantom{\dagger}}_{k+n})\,
{{\bar\theta}}^{\vphantom{\dagger}}_k\,{{\bar\theta}}^{\vphantom{\dagger}}_{k+n}\ .\nonumber
\end{align}
The correlation function may then be written
\begin{equation}
C^{\vphantom{\dagger}}_{{\scriptscriptstyle{\rm SS}}}(n)={1\over\sqrt{2}}(2M+1)^2\cdot{{\cal N}\over{\cal D}}\ .
\end{equation}

\begin{figure}[!t]
\centering
\includegraphics[width=8cm,height=7cm]{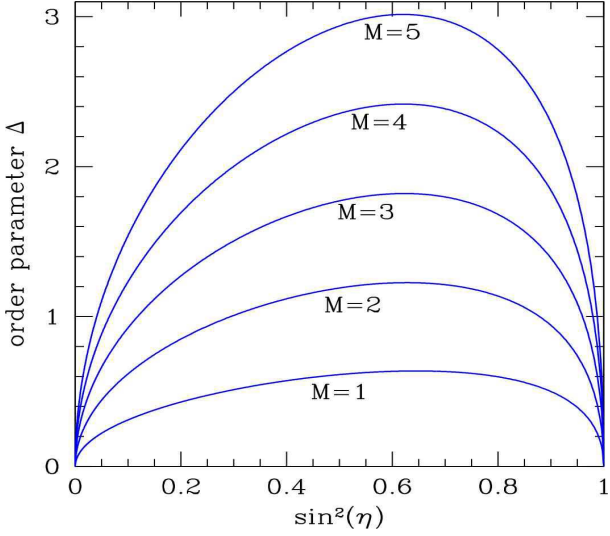}
\caption
{\label{opar} Order parameter $\Delta=\langle(a_n\,b_{n+1} - b_n\,a_{n+1})
\,f^{\dagger}_n f^{\dagger}_{n+1}\rangle$ in the general SVBS chain.  The parameter $\eta$
is given by $\eta=\tan^{-1} |r|$.}
\end{figure}

The calculation of the numerator ${\cal N}$ proceeds along the same lines
as that of ${\cal D}$.  Starting with site $0$, we generate the expression
$\alpha^{\vphantom{\dagger}}_j + \beta^{\vphantom{\dagger}}_n\theta^{\vphantom{\dagger}}_j{{\bar\theta}}^{\vphantom{\dagger}}_j$.   When we arrive at site $k$,
only the second term on the RHS of Eq.(\ref{expand}) contributes.  We then have
\begin{align}
&\alpha^{\vphantom{\dagger}}_k\!\int\!\!{d{\hat{\mib n}}^{\vphantom{\dagger}}_k\over 4\pi}\!\!\int\!\!d{{\bar\theta}}^{\vphantom{\dagger}}_k\!\!
\int\!\!d\theta^{\vphantom{\dagger}}_k\,e^{(2M+1)\theta^{\vphantom{\dagger}}_k{{\bar\theta}}^{\vphantom{\dagger}}_k}\\
&\qquad\qquad\times(u^{\vphantom{\dagger}}_k\,v^{\vphantom{\dagger}}_{k+n}-v^{\vphantom{\dagger}}_k\,u^{\vphantom{\dagger}}_{k+n})\,{{\bar\theta}}^{\vphantom{\dagger}}_k\,{{\bar\theta}}^{\vphantom{\dagger}}_{k+n}\nonumber\\
&\qquad\qquad\times\big|u^{\vphantom{\dagger}}_k\,v^{\vphantom{\dagger}}_{k+1} - v^{\vphantom{\dagger}}_k\,u^{\vphantom{\dagger}}_{k+1}+r\,\theta^{\vphantom{\dagger}}_k\theta^{\vphantom{\dagger}}_{k+1}\big|^{2M}
{\vphantom{\sum_N^N}}\nonumber\\
&=-\alpha^{\vphantom{\dagger}}_k\,Mr\!\int\!\!{d{\hat{\mib n}}^{\vphantom{\dagger}}_k\over 4\pi}\,\bigg(
{1-{\hat{\mib n}}^{\vphantom{\dagger}}_k\cdot{\hat{\mib n}}_{k+1}\over 2}\bigg)^{\!\!M-1}\!\!\!
({{\bar u}}^{\vphantom{\dagger}}_k\,{{\bar v}}^{\vphantom{\dagger}}_{k+1}-{{\bar v}}^{\vphantom{\dagger}}_k\,{{\bar u}}^{\vphantom{\dagger}}_{k+1})\nonumber\\
&\qquad\qquad\times (u^{\vphantom{\dagger}}_k\,v^{\vphantom{\dagger}}_{k+n}-v^{\vphantom{\dagger}}_k\,u^{\vphantom{\dagger}}_{k+n})\,
\,\theta^{\vphantom{\dagger}}_{k+1}\,{{\bar\theta}}^{\vphantom{\dagger}}_{k+n}{\vphantom{\sum_i}}
\nonumber\\
&=-\bigg({M\,r\over M+1}\bigg)\,\alpha^{\vphantom{\dagger}}_k\,({{\bar u}}^{\vphantom{\dagger}}_{k+1}\,u^{\vphantom{\dagger}}_{k+n}+
{{\bar v}}^{\vphantom{\dagger}}_{k+1}\,v^{\vphantom{\dagger}}_{k+n})\,\theta^{\vphantom{\dagger}}_{k+1}\,{{\bar\theta}}^{\vphantom{\dagger}}_{k+n}\ .\nonumber
\end{align}
When we integrate over site $k+1$, we obtain
\begin{equation}
\bigg({M\,|r|\over M+1}\bigg)^{\!\!2}\,\alpha^{\vphantom{\dagger}}_k\,(u^{\vphantom{\dagger}}_{k+2}\,v^{\vphantom{\dagger}}_{k+n}-
v^{\vphantom{\dagger}}_{k+2}\,u^{\vphantom{\dagger}}_{k+n})\,{{\bar\theta}}^{\vphantom{\dagger}}_{k+2}\,{{\bar\theta}}^{\vphantom{\dagger}}_{k+n}\ .
\end{equation}
We have now replicated the form of the integrand.  Clearly whenever
$n$ is even, the numerator ${\cal N} $ vanishes.  For $n$ odd, we obtain
\begin{equation}
\beta^{\vphantom{\dagger}}_{k+n}=\bigg({M\,|r|\over M+1}\bigg)^{\!\!n}\,
\delta^{\vphantom{\dagger}}_{n,{\rm odd}}\cdot\alpha^{\vphantom{\dagger}}_k\ .
\end{equation}
The  correlation length $\zeta(M,r)$ is then given by
\begin{align}
e^{-1/\zeta(M,r)}&={1\over\lambda_+}\,\bigg({M\,|r|\over M+1}\bigg)\\
&={M\,|r|\over M+{\frac{1}{2}}+\sqrt{(M+{\frac{1}{2}})^2 + M(M+1)|r|^2}}\ ,\nonumber
\end{align}
where
\begin{equation}
\lambda^{\vphantom{\dagger}}_+={M+{1\over 2}\over M+1} + \sqrt{\left({M+{1\over 2}\over M+1} \right)^{\!\!2} + {M|r|^2\over M+1}}
\end{equation}
is the largest eigenvalue of the matrix ${\cal D} $ from Eq.(\ref{deqn}).
We can define the $s$-wave order parameter as
\begin{align}
{\rm{\Delta}}&=\langle \,(a^{\vphantom{\dagger}}_k\,b^{\vphantom{\dagger}}_{k+1} - b^{\vphantom{\dagger}}_k\,a^{\vphantom{\dagger}}_{k+1})\,f^{\dagger}_k\,f^{\dagger}_{k+1}\,\rangle\nonumber\\
&={2M(M+{\frac{1}{2}})^2\,r\over \left(\sqrt{M(M\!+\!1)(1\!+\!|r|^2) + {1\over 4}}+{1\over 2}(M\!+\!{1\over 2})\right)^{\!2}-
{1\over 4}(M\!+\!{1\over 2})^2}\ .
\label{swaveorderpara}
\end{align}
($\Delta$ is plotted in Fig. \ref{opar}.) 

\subsubsection{Spin correlations}
We next turn to the spin-spin correlation function,
$C^{\vphantom{\dagger}}_{{\scriptscriptstyle{\rm
spin}}}(n)=\big\langle \,{\mib L} (j)\cdot{\mib L} (j+n) \,
\big\rangle$. The spin operator is given by ${\mib L} ={\frac{1}{2}}
b^{\dagger}_\mu{\mib\sigma}_{\mu\nu}b^{\vphantom{\dagger}}_\nu$, and is of the type
${\hat T}^0_{k=1}$. Accordingly, Eq.(\ref{repa}) gives the prescription\begin{equation}
{\mib L} (j)\to (M+{\frac{1}{2}})\,{\hat{\mib n}}^{\vphantom{\dagger}}_j\,e^{\theta^{\vphantom{\dagger}}_j{{\bar\theta}}^{\vphantom{\dagger}}_j}\ .
\end{equation}
Once again, the correlation function is expressed as a ratio of ${\cal N}/{\cal D}$.  In the numerator, when
we arrive at site $k$, we have the integral
\begin{align}
&(M+{\frac{1}{2}})\!\int\!\!{d{\hat{\mib n}}^{\vphantom{\dagger}}_k\over 4\pi}\!\!\int\!\!d{{\bar\theta}}^{\vphantom{\dagger}}_k\!\!\int\!\!
d\theta^{\vphantom{\dagger}}_k\, e^{2(M+1)\theta_k{{\bar\theta}}_k}\nonumber\\
&\qquad\times\big(\alpha^{\vphantom{\dagger}}_k+\beta^{\vphantom{\dagger}}_k\theta^{\vphantom{\dagger}}_k
{{\bar\theta}}^{\vphantom{\dagger}}_k\big)\,{\hat{\mib n}}^{\vphantom{\dagger}}_k\,\Bigg[\bigg({1-{\hat{\mib n}}^{\vphantom{\dagger}}_k\cdot{\hat{\mib n}}^{\vphantom{\dagger}}_{k+1}\over 2}
\bigg)^{\!\!M}\nonumber\\
& \qquad+ M^2 |r|^2 \bigg({1-{\hat{\mib n}}^{\vphantom{\dagger}}_k\cdot{\hat{\mib n}}^{\vphantom{\dagger}}_{k+1}
\over 2}\bigg)^{\!\!M-1}\theta^{\vphantom{\dagger}}_k{{\bar\theta}}^{\vphantom{\dagger}}_k\,\theta^{\vphantom{\dagger}}_{k+1}{{\bar\theta}}^{\vphantom{\dagger}}_{k+1}\Bigg]\nonumber\\
&\qquad=\big(\alpha^{\vphantom{\dagger}}_{k+1}+\beta^{\vphantom{\dagger}}_{k+1}\theta^{\vphantom{\dagger}}_{k+1}{{\bar\theta}}^{\vphantom{\dagger}}_{k+1}\big)\,
{\hat{\mib n}}^{\vphantom{\dagger}}_{k+1}\ ,
\end{align}
with
\begin{equation}
\begin{pmatrix}\alpha^{\vphantom{\dagger}}_{k+1} \\ \\ \beta^{\vphantom{\dagger}}_{k+1}\end{pmatrix}=\ \
{M(M\! + \!{\frac{1}{2}})\over M\! + \!1}\begin{pmatrix}
{2(M+1)\over M+2} && {1\over M+2} \\ && \\
(M-1)|r|^2 && 0 \end{pmatrix}
\begin{pmatrix}\alpha^{\vphantom{\dagger}}_k \\ \\ \beta^{\vphantom{\dagger}}_k\end{pmatrix}\ .
\end{equation}
For sites $j$ between $k$ and $k+n$, we have
\begin{align}
&\int\!\!{d{\hat{\mib n}}^{\vphantom{\dagger}}_j\over 4\pi}\!\!\int\!\!d{{\bar\theta}}^{\vphantom{\dagger}}_j\!\!\int\!\!
d\theta^{\vphantom{\dagger}}_j\, e^{(2M+1)\theta_j{{\bar\theta}}_j}\nonumber\\
&\qquad\times\big(\alpha^{\vphantom{\dagger}}_j+\beta^{\vphantom{\dagger}}_j\theta^{\vphantom{\dagger}}_j
{{\bar\theta}}^{\vphantom{\dagger}}_j\big)\,{\hat{\mib n}}^{\vphantom{\dagger}}_j\,\Bigg[\bigg({1-{\hat{\mib n}}^{\vphantom{\dagger}}_j\cdot{\hat{\mib n}}^{\vphantom{\dagger}}_{j+1}\over 2}
\bigg)^{\!\!M}\nonumber\\
& \qquad+ M^2 |r|^2 \bigg({1-{\hat{\mib n}}^{\vphantom{\dagger}}_j\cdot{\hat{\mib n}}^{\vphantom{\dagger}}_{j+1}
\over 2}\bigg)^{\!\!M-1}\theta^{\vphantom{\dagger}}_j{{\bar\theta}}^{\vphantom{\dagger}}_j\,\theta^{\vphantom{\dagger}}_{j+1}{{\bar\theta}}^{\vphantom{\dagger}}_{j+1}\Bigg]\nonumber\\
&\qquad=\big(\alpha^{\vphantom{\dagger}}_{j+1}+\beta^{\vphantom{\dagger}}_{j+1}\theta^{\vphantom{\dagger}}_{j+1}{{\bar\theta}}^{\vphantom{\dagger}}_{j+1}\big)\,
{\hat{\mib n}}^{\vphantom{\dagger}}_{j+1}\ ,
\end{align}
with
\begin{equation}
\begin{pmatrix}\alpha^{\vphantom{\dagger}}_{j+1} \\ \\ \beta^{\vphantom{\dagger}}_{j+1}\end{pmatrix}={\cal K}
\begin{pmatrix}\alpha^{\vphantom{\dagger}}_j \\ \\ \beta^{\vphantom{\dagger}}_j\end{pmatrix}\ ,
\end{equation}
where
\begin{equation}
{\cal K} =-\,{M\over (M+1)(M+2)}\begin{pmatrix}
2M+1 && 1 \\ && \\ (M-1)(M+2)|r|^2 && 0 \end{pmatrix}\ .
\end{equation}
Finally, we come to site $k+n$, where we have ${\hat{\mib n}}^{\vphantom{\dagger}}_{k+n}\cdot{\hat{\mib n}}^{\vphantom{\dagger}}_{k+n}=1$,
and
\begin{align}
&(M+{\frac{1}{2}})\!\int\!\!{d{\hat{\mib n}}^{\vphantom{\dagger}}_{k+n}\over 4\pi}\!\!\int\!\!d{{\bar\theta}}^{\vphantom{\dagger}}_{k+n}\!\!\int\!\!
d\theta^{\vphantom{\dagger}}_{k+n}\, e^{2(M+1)\theta_{k+n}{{\bar\theta}}_{k+n}}\nonumber\\
&\qquad\times\big(\alpha^{\vphantom{\dagger}}_{k+n}+\beta^{\vphantom{\dagger}}_{k+n}\theta^{\vphantom{\dagger}}_{k+n}
{{\bar\theta}}^{\vphantom{\dagger}}_{k+n}\big)\,\Bigg[\bigg({1-{\hat{\mib n}}^{\vphantom{\dagger}}_{k+n}\cdot{\hat{\mib n}}^{\vphantom{\dagger}}_{k+n+1}\over 2}
\bigg)^{\!\!M}\nonumber\\
& \qquad\qquad+ M^2 |r|^2 \bigg({1-{\hat{\mib n}}^{\vphantom{\dagger}}_{k+n}\cdot{\hat{\mib n}}^{\vphantom{\dagger}}_{k+n+1}
\over 2}\bigg)^{\!\!M-1}\nonumber\\
&\qquad\qquad\qquad\times \theta^{\vphantom{\dagger}}_{k+n}{{\bar\theta}}^{\vphantom{\dagger}}_{k+n}\,\theta^{\vphantom{\dagger}}_{k+n+1}{{\bar\theta}}^{\vphantom{\dagger}}_{k+n+1}\Bigg]\nonumber\\
&\qquad=\big(\alpha^{\vphantom{\dagger}}_{k+n+1}+\beta^{\vphantom{\dagger}}_{k+n+1}\theta^{\vphantom{\dagger}}_{k+n+1}{{\bar\theta}}^{\vphantom{\dagger}}_{k+n+1}\big)\ ,
\end{align}
with
\begin{equation}
\begin{pmatrix}\alpha^{\vphantom{\dagger}}_{k+n+1} \\ \\ \beta^{\vphantom{\dagger}}_{k+n+1}\end{pmatrix}=\ \
(M+{\frac{1}{2}})\begin{pmatrix}
2 && {1\over M+1} \\ && \\ M|r|^2 && 0 \end{pmatrix} 
\begin{pmatrix}\alpha^{\vphantom{\dagger}}_{k+n} \\ \\ \beta^{\vphantom{\dagger}}_{k+n}\end{pmatrix}\ .
\end{equation}
For sites $l>k+n$, we propagate by the matrix ${\cal D} $ from the denominator.

Assuming $N\to\infty$, with $n$ finite but large, we can ignore the ends, and we obtain
\begin{align}
C^{\vphantom{\dagger}}_{{\scriptscriptstyle{\rm spin}}}(n)&=A\big(\lambda^{\vphantom{\dagger}}_{\cal K} /\lambda^{\vphantom{\dagger}}_{\cal D} \big)^{|n|}\nonumber\\
&=A\,(-1)^n\,e^{-|n|/\xi(M,r)}\ ,
\end{align}
where $A$ is a coefficient, and $\lambda^{\vphantom{\dagger}}_{{\cal K} ,{\cal D} }$ are the largest magnitude eigenvalues
of the matrices ${\cal K} $ and ${\cal D} $, respectively.  The spin correlation length is thus given by
\begin{align}
e^{-1/\xi(M,r)}&=-\,{\lambda^{\vphantom{\dagger}}_{\cal K}\over\lambda^{\vphantom{\dagger}}_{\cal D} }\\
&={M\over M+2}\cdot\left( {1+\sqrt{1+ {(M-1)(M+2)|r|^2\over (M+{1\over 2})^2}}\over
1+\sqrt{1 + {M(M+1)|r|^2\over (M+{1\over 2})^2} } }\right)\ .
\nonumber
\end{align}

\begin{figure}[!t]
\centering
\includegraphics[width=8cm,height=7cm]{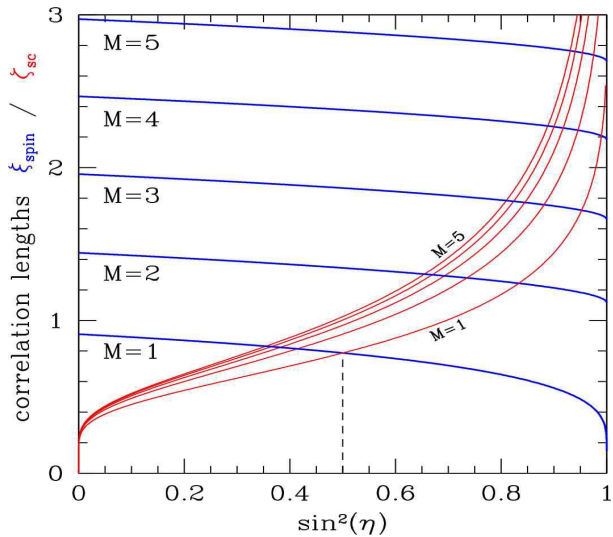}
\caption
{\label{corr} Spin correlation length $\xi(M,r)$ (blue) and superconducting correlation
length $\zeta(M,r)$ (red) for the general SUSY AKLT chain.}
\end{figure}

In the $r\to 0$ limit we recover the result $C(n)=A\,\big(\!-\frac{M}{M+2}\big)^{|n|}$ found for general AKLT chains in Ref. \cite{arovas1988fit}. 
Note that for $r\to\infty$ and $M=1$ the correlation length vanishes.
This is because in this limit the ground state is that for the $S={\frac{1}{2}}$ Majumdar-Ghosh
model, {\it i.e.}, alternating singlets, for which there are no correlations beyond
nearest neighbors.  For the $M>1$ generalizations of Majumdar-Ghosh, however,
the correlation length is finite.  The spin correlation length $\xi(M,r)$ and superconducting
correlation length $\zeta(M,r)$ are both plotted in Fig. \ref{corr}, {\it versus\/} the
parameter $\sin^2\!\eta\equiv |r|^2/(1+|r|^2)$.
These two correlation lengths coincide at $r=(2M+1)/3$.
Especially, when $M=1$, they coincide at $r=1$.

\section{Relation to QHE States}\label{QHE}

Here, we  discuss analogies between the lowest Landau level (LLL) physics and the spin physics in detail.  Much of our discussion is an extension of the pioneering work by Haldane
\cite{haldane1983fqh} on the FQHE in a spherical geometry.

\begin{table}
\renewcommand{\arraystretch}{1.5}
\begin{center}
\begin{tabular}{|c||c|c|}
\hline      & LLL physics &   Spin physics  \\
\hline
\hline  Space  & External    & Internal
\\
\hline   Quantum number  &  Monopole charge   & Spin magnitude
\\ \hline    Basic state &  Hopf spinor  & Spin-coherent state
 \\  \hline
 Manifold & Fuzzy sphere &  Bloch sphere
\\ \hline
\end{tabular}
\end{center}
\caption{Correspondences between LLL physics and  spin physics}
\label{correspondence}
\end{table}

We begin with a discussion about analogies in one-particle problem.  For
The LLL bases are given by the monopole harmonics \cite{wuyang1976}, which form an irreducible representation of $\textsf{SU}(2)$ indexed by the unique Casimir operator, which is the monopole charge.  As is well known,
the monopole harmonics in the LLL are constructed as symmetric products of Hopf spinors. Mathematically, the Hopf spinor is equivalent to a spin coherent state for a state in the fundamental
($S=\frac{1}{2}$) representation.
The symmetric products of the spin coherent states give rise to higher spin states.  In the LLL, the kinetic term is quenched and the coordinates of the two-spheres are effectively reduced to operators of $\textsf{SU}(2)$ algebra. Such manifold with noncommutative coordinates is known as the fuzzy sphere and its mathematical structure is equal to the Bloch sphere  of spin physics. The relations between the LLL states and the spin states are summarized in Table \ref{correspondence}.
Thus, as for the one-particle problem, there are apparent analogies  between the LLL physics and the spin physics.

\subsection{Laughlin-Haldane and  AKLT states }

Even in many-body level, as briefly mentioned in Sec.\ref{sectintro}, remarkable resemblances between the Laughlin state
and the AKLT state have been reported in the work of one of the
authors \cite{arovas1988ehm}. On Haldane's two-spheres, particles are
uniformly distributed to form a rotationally invariant
incompressible liquid described by the Laughlin-Haldane function, 
\begin{equation}
\Phi_{\rm LH}^{(m)}=\prod_{i<j}^N(u_iv_j-v_iu_j)^m,
\label{Llinori}
\end{equation}
where $(u,v)$ indicates the Hopf spinor. Meanwhile, the AKLT state is the VBS state made by the $\textsf{SU}(2)$ singlet combination of Schwinger bosons [Eq.(\ref{akltstateoperator})], and, in the spin-coherent state representation, is written as
\begin{equation}
\Phi_{\rm AKLT}^{(M)}=\prod_{\langle ij\rangle}^z (u_i v_j-v_i u_j)^M.
\label{AKLTori}
\end{equation}
Obvious resemblances may be found between Eqs.(\ref{Llinori}) and (\ref{AKLTori}). The power $m$ in the Laughlin-Haldane state takes even or odd integer depending on the statistics of the particles, while $M$ in the AKLT state specifies the number of the valence bonds on each site and has nothing to do with statistics.
Since these two states are ``almost'' mathematically equivalent, their truncated pseudopotential Hamiltonians are similarly constructed by the form of two-body interactions: the truncated Hamiltonian for the AKLT state is given by Eq.(\ref{AKLTHamiltonian}), while for the Laughlin-Haldane state, it is given by
\begin{equation}
H=\sum_{i<j}\sum_{J^*+1}^{2S} V_J P_J(i,j),
\end{equation}
where $J^*=2S-m$ with $S=m(N-1)/2$.
Based on the  $\textsf{OSp}(1|2)$ supergroup analysis, the SUSY Laughlin-Haldane wave function was proposed as
\begin{equation}
\Psi_{\rm SLH}^{(m)}=\prod_{i<j}^N(u_iv_j-v_iu_j+r\theta_i\theta_j)^m,
\label{parameterSUSYLlin}
\end{equation}
where $(u,v,\theta)$ indicates the SUSY Hopf spinor.
In Ref.{\cite{hasebe2005PRL}}, $r$ is fixed as $-1$, but here we take $r$ as a free parameter.
Extracting the original Laughlin-Haldane wave function, the SUSY Laughlin-Haldane  state can be rewritten as
\begin{equation}
\Psi_{\rm SLH}^{(m)}=\exp\biggl(mr\sum_{i<j}\frac{\theta_i\theta_j}{u_iv_j-v_iu_j}\biggr)\cdot\Phi_{\rm LH}^{(m)}.
\label{exponentSUSYLlin}
\end{equation}
All of the important physics are included in the exponential factor of Eq.(\ref{exponentSUSYLlin}), and
this deformation enables us to perform an intuitive interpretation  of the SUSY Laughlin-Haldane wave function. The denominator of the exponential factor, $1/(u_iv_j-v_iu_j)$, represents a $p$-wave bound state of two particles $i$ and $j$, and the SUSY Laughlin state is regarded as a $p$-wave superfluid on the original Laughlin state \cite{hasebe2008ula}.
By expanding the exponent, one may find
\begin{align}
&\Psi_{\rm SLH}^{(m)}= \Phi_{\rm LH}^{(m)}+mr\biggl(\sum_{i<j}^N\frac{\theta_i\theta_j}{u_iv_j-v_iu_j}\biggr) \cdot\Phi_{\rm LH}^{(m)} \nonumber\\
&~~~+\frac{1}{2}(mr)^2\biggl(\sum_{i<j}^N\frac{\theta_i\theta_j}{u_iv_j-v_iu_j}\biggr)^2\cdot\Phi_{\rm LH}^{(m)} +\cdots  \nonumber\\
& ~~~ +(mr)^{N/2} \prod_i^N \theta_i\cdot {\cal A}\Bigg(\prod_{j:{\rm even}} \frac{1}{u_{j-1}v_j-v_{j-1}u_j  }\Bigg) \cdot\Phi_{\rm LH}^{(m)},
\label{expLlin1}
\end{align}
where ${\cal A}$ in the last term represents antisymmetrization over all different choices of breaking particles into pairs, and is simply known as the Pfaffian. Hence, the last term in Eq.(\ref{expLlin1}) represents the Pfaffian state proposed by Moore and Read \cite{moore1991nfq}
\begin{align}
&\Phi_{\rm MR}^{(m)}= {\cal A}\prod_{i:\rm{even}} \frac{1}{u_{i-1}v_i-v_{i-1}u_i  } \cdot\Phi_{\rm LH}^{(m)}\nonumber\\
&~~~~~~=     {\rm Pf}\>\bigg(\frac{1}{u_iv_j-v_iu_j}\bigg)\cdot\Phi_{\rm LH}^{(m)},
\end{align}
where all of the particles form $p$-wave pairings to form a bosonic QH state.
It is noted that the expression (\ref{expLlin1}) should be regarded as the expansion about the parameter $r$ not $m$, since the original Laughlin-Haldane function itself depends on $m$.

\subsection{Physical interpretation of the SVBS state}\label{physicalintSVBS}

\begin{figure}[!t]
\centering
\includegraphics[width=7.5cm]{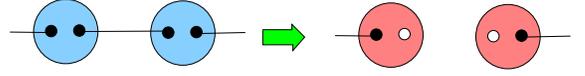}
\caption
{\label{bondbreakingfig} The single-bond breaking operator annihilates a valence bond  and creates a fermion pair on the nearest-neighbor sites.}
\end{figure}

\begin{figure}[!t]
\centering
\includegraphics[width=7.5cm]{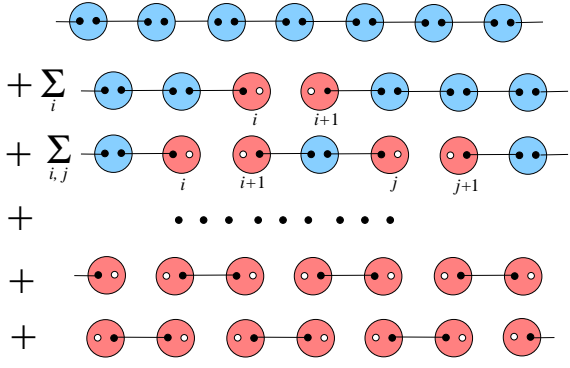}
\caption
{\label{expSUSYAKLTfig} Graphical representation for the expansion of the SUSY AKLT spin state with $M=1$ [Eq.(\ref{expAKLT1})].
At the $n^{\rm th}$ term of the expansion, there appears the superposition of AKLT states with $(n-1)$ hole pairs.
In particular, the original AKLT state is realized as the first term ($r\rightarrow 0$) and the MG dimer states are realized as the last term
($r\rightarrow \infty$).}
\end{figure}

Inspired by the similarity between the Laughlin-Haldane and the AKLT states, from the SUSY Laughlin-Haldane wave function [(Eq.(\ref{parameterSUSYLlin})], one may derive the SUSY AKLT state, 
\begin{equation}
\Psi_{\rm AKLT}^{(M)} =\prod_{\langle ij\rangle}^z(u_iv_j-v_iu_j+r\eta_i\eta_j)^M,
\end{equation}
which is the spin-hole coherent representation of Eq.(\ref{SAKLT}).
 In the following, we focus on the SVBS spin chain. Just as in the SUSY Laughlin-Haldane case, the SVBS spin chain state $z=2$ is rewritten as
\begin{equation}
\Psi_{\rm AKLT}^{(M)}=\exp\biggl(Mr\sum_{i}\frac{\theta_i\theta_{i+1}}{u_iv_{i+1}-v_iu_{i+1}}\biggr)\cdot\Phi_{\rm AKLT}^{(M)},
\end{equation}
where the exponential factor $\theta_i\theta_j/(u_iv_{i+1}-v_iu_{i+1})$ which we call the ``pair creator'' has the following physical interpretation: it replaces one of the valence
bonds between sites $i$ and $i+1$ by a fermion (hole) pair; this is depicted in
Fig. \ref{bondbreakingfig}.
The SVBS chain state is expanded as
\begin{align}
&\Psi_{\rm AKLT}^{(M)}\nonumber\\
&~~~= \Phi_{\rm AKLT}^{(M)}+Mr\biggl(\sum_{i}\frac{\theta_i\theta_{i+1}}{u_iv_{i+1}-v_iu_{i+1}}\biggr) \cdot\Phi_{\rm AKLT}^{(M)}\nonumber\\
&~~~+\frac{1}{2}(Mr)^2\biggl(\sum_{i}\frac{\theta_i\theta_{i+1}}{u_iv_{i+1}-v_iu_{i+1}}\biggr)^2\cdot\Phi_{\rm AKLT}^{(M)}+\cdots\nonumber\\
& ~~~  +(Mr)^{L/2}\prod_j \theta_j \left(\prod_{i\atop \rm even}-\prod_{i\atop \rm odd}\right) \frac{1}{u_i v_{i+1}-v_i u_{i+1}}\cdot \Phi_{\rm AKLT}^{(M)}.
\label{expAKLT1}
\end{align}
We assume here that the total number of sites $L$ in our ring is even.
The original AKLT state appears as the first term in this expansion in powers of
the Grassmann coordinates.  The second term consists of superpositions of all
AKLT states with one hole pair, the third term of all superpositions with two hole pairs,
{\it etc.\/}  The final term in the expansion contains the product $\theta_1\cdots\theta_L$
over all sites.  Its corresponding spin wave function is a superposition of two
generalized Majumdar-Ghosh states, one in which a valence bond has been removed
from each even link $(2n,2n+1)$, and the other where a valence bond has been
removed from each odd link $(2n-1,2n)$.  Note that each site can accommodate at
most one hole (Fig. \ref{expSUSYAKLTfig}). 
As discussed in \S \ref{sectintro}, for $M=1$ the last term of Eq.(\ref{expAKLT1}) gives
precisely the $S=\frac{1}{2}$ Majumdar-Ghosh state,
\begin{align}
&\left(\prod_{i\atop\rm even}-\prod_{i\atop\rm odd}\right)
\frac{1}{u_i v_{i+1}-v_i u_{i+1}}\cdot \Phi_{\rm AKLT}^{(M=1)}\nonumber\\
&\qquad\qquad= \left(\prod_{i\atop\rm even}-\prod_{i\atop\rm odd}\right)
(u_i v_{i+1}-v_i u_{i+1})\nonumber\\
&\qquad\qquad=\Phi_{{\scriptscriptstyle{\rm A}}}-\Phi_{{\scriptscriptstyle{\rm B}}},
\end{align}
where $\Phi_{{\scriptscriptstyle{\rm A}}}$ and $\Phi_{{\scriptscriptstyle{\rm B}}}$ correspond to the two dimer states of Eq.(\ref{twodegMGstates}).

Thus, both the Majumdar-Ghosh and Moore-Read states appear as the last terms in the expansion of the corresponding super wave functions.  It is interesting to note in this
regard that both the Majumdar-Ghosh and Moore-Read wave functions vansh when any three particles (Moore-Read)
or any three neighboring spins (Majumdar-Ghosh) coincide, and their truncated pseudopotential
Hamiltonians are constructed by three-body interactions \cite{greiter1991phs}.
For the Moore-Read state,
\begin{equation}
H_{\rm MR}=\sum_{i<j<k}\sum_{J=3(S-m)+2}^{3S} V_J\, P_J(i,j,k)
\end{equation}
with $S=\frac{1}{2} [m(N-1)-1]$, while for the Majumdar-Ghosh state
\begin{equation}
H_{\rm MG}=V_{{3\over 2}}\sum_{i}  P_{{3\over 2}}(i,i+1,i+2).
\end{equation}

\subsection{More fermion coordinates}
\begin{figure}[!t]
\centering
\includegraphics[width=7.5cm]{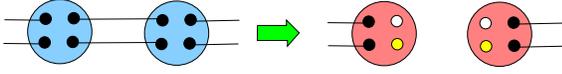}
\caption
{\label{bondbreaking2fig} The operation of the double-bond breaking operator.
The white circles represent the hole pair of $\theta_i\theta_{i+1}$, while the light yellow circles represent the other hole pair of $\eta_i\eta_{i+1}$.}
\end{figure}

\begin{figure}[!t]
\centering
\includegraphics[width=7.5cm]{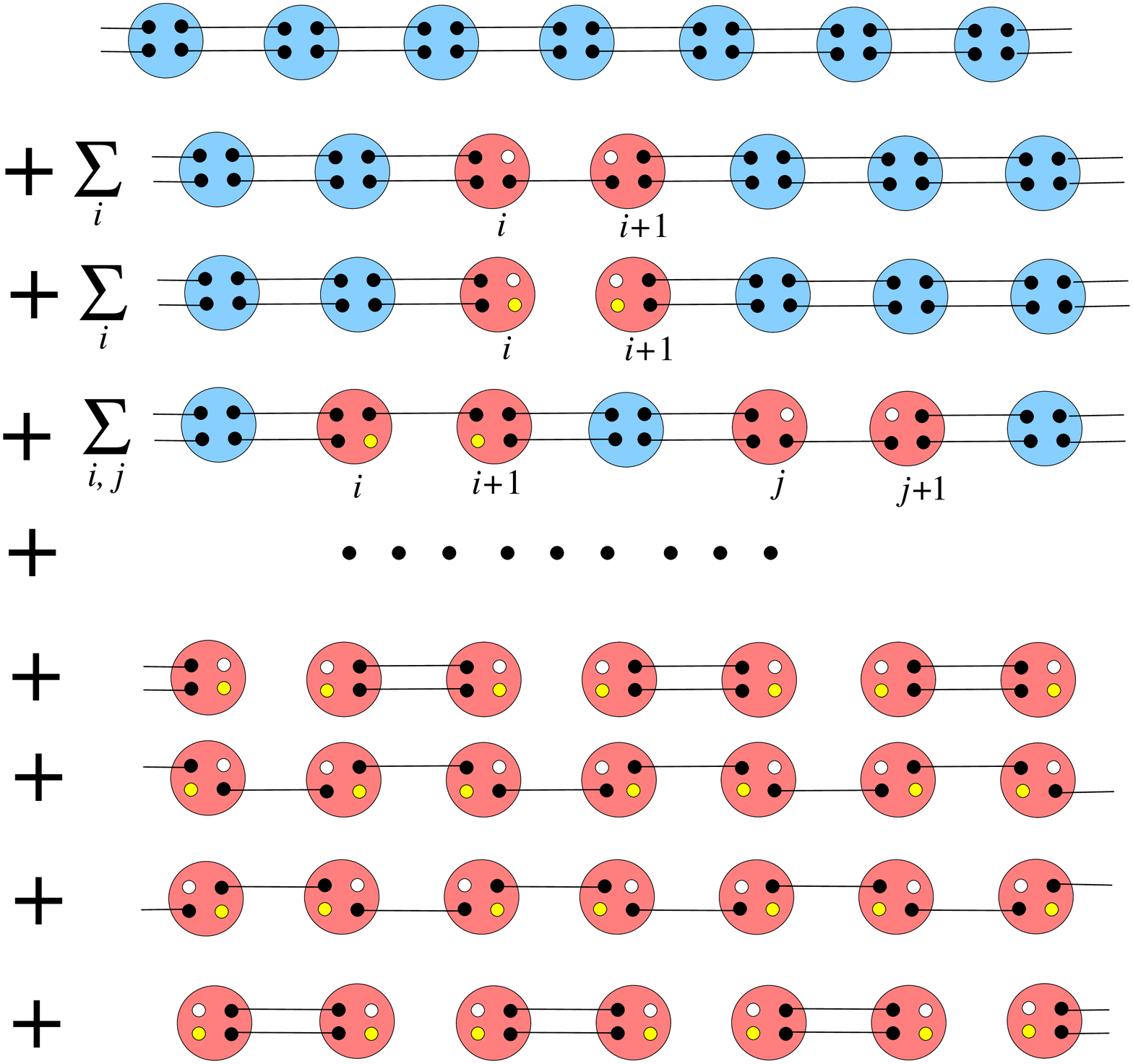}
\caption
{\label{expSUSYAKLT2fig} The graphical representation for the expansion of the generalized AKLT spin chain state [Eq.(\ref{extraAKLT2})]. The first term represents the original $M=2$ AKLT state. At the second term, the superposition of the AKLT states with one hole pair appears. At both third and fourth terms, one may find the AKLT states with two hole pairs. At the third term, the two holes are generated by the double-bond breaking operation, while at the fourth term, they are generated by two successive different single-bond breaking operations. At the last terms of the expansion, we obtain four states two of which are fully dimerized states, and the other two are partially dimerized states that are equal to the $M=1$ AKLT states.}
\end{figure}

Our construction may be generalized to include additional Grassmann coordinates.
Introducing two Grassmann species $\theta_i$ and $\eta_i$,  we write the extended SUSY Laughlin-Haldane wave function as
\begin{equation}
\Psi_{\rm SLH}^{(m)} =\prod_{i<j}(u_iv_j-v_i u_j+r_1\,\theta_i\theta_j+r_2\,\eta_i\eta_j)^m,
\end{equation}
where $r_1$ and $r_2$ are two free parameters. We may now write
\begin{align}
\Psi_{\rm SLH}^{(m)}&=\exp\!\bigg(mr_1\sum_{i<j}\frac{\theta_i\theta_j}{u_iv_j-v_iu_j}\bigg)
\label{exponentsoftwofermion}\\
&\quad\cdot\exp\bigg(mr_2\sum_{i<j}\frac{\eta_i\eta_j}{u_iv_j-v_iu_j}\bigg)\nonumber\\
&\quad \cdot \exp\bigg(-mr_1r_2\sum_{i<j}\frac{\theta_i\theta_j\eta_i\eta_j}{(u_i v_j-v_i u_j)^2}\bigg)\cdot \Phi_{\rm SLH}^{(m)}\ .\nonumber
\end{align}
We have already encountered the first and second exponents of Eq.(\ref{exponentsoftwofermion}) in the previous analysis, each of which represents the $p$-wave pairing state.
The third exponent is the newly appeared term, and its exponential factor provides $(-1)^2$ by the interchange of $i$ and $j$ to suggest the property of $d$-wave pairing.
When we expand the third exponent, at the last term, we obtain
\begin{align}
& \int \prod_i^N d\theta_i  d\eta_i\cdot  \exp\biggl(-mr_1r_2\sum_{i<j}\frac{\theta_i\theta_j\eta_i\eta_j}{(u_i v_j-v_i u_j)^2}\biggr)\nonumber\\
&   \qquad\qquad=(-mr_1r_2)^{N/2} \,    S \biggl(\frac{1}{(u_iv_j-v_iu_j)^2}\biggr)\ ,
\label{lasttwofermionllin}
\end{align}
where $\prod_i d\theta_i d\eta_i\equiv \prod_i d\theta_i\prod_i d \eta_i$, and $S$ represents the symmetrization operation, which is realized by  changing all the signs of terms in Pfaffian to be plus, and is known to yield
the Haffnian,
\begin{equation}
S \biggl(
\frac{1}{(u_iv_j-v_iu_j)^2}\biggr) ={\rm Hf} \biggl(
\frac{1}{(u_iv_j-v_iu_j)^2}\biggr).
\end{equation}
The first and second exponents in Eq.(\ref{exponentsoftwofermion}) are expanded as in Eq.(\ref{expLlin1}) to yield the product of two Pfaffians, and  produce the Haffnian again,
\begin{align}
&(mr_1)^{N/2}(m r_2)^{N/2}\, {\rm Pf}^2\biggl(
\frac{1}{(u_iv_j-v_iu_j)^2}\biggr)\nonumber\\
&~~~~~~~~~~~~~~~=m^N (r_1r_2)^{N/2}\, {\rm Hf}\biggl(
\frac{1}{(u_iv_j-v_iu_j)^2}\biggr).
\end{align}
Besides this, there are many cross terms to yield Haffnian in the products of expansions of the three exponents.
Collecting all of the contributions,
the last term of the expansion [Eq.(\ref{exponentsoftwofermion})] is summarized as
\begin{equation}
\int \!\!\prod_i d\theta_i d\eta_i \Psi_{\rm SLH}^{(m)}
= \big(m(m-1)\,r_1r_2\big)^{N/2}\cdot \Phi_{\rm HR}^{(m)},\label{appearanceofhr}
\end{equation}
where $\Phi_{\rm HR}$ is the Haffnian state of Haldane-Rezayi \cite{haldane1988ssw}, 
\begin{equation}
\Phi_{\rm HR}^{(m)}= {\rm Hf}\biggl(\!\!
\frac{1}{(u_iv_j-v_iu_j)^2}\!\!\biggr)\cdot\Phi_{\rm SLH}^{(m)},
\end{equation}
which represents $d$-wave pairing QH state.
We see that the Laughlin wave function with two Grassmann species [Eq.(\ref{exponentsoftwofermion})] is expanded as
\begin{align}
\Psi_{\text{SLH}}^{(m)}&=\Phi_{\text{LH}}^{(m)}+\ldots\nonumber\\
&\qquad+\bigg((mr_1)^{N/2}\prod_i \theta_i+ (mr_2)^{N/2}\prod_i \eta_i\bigg)
\cdot\Phi_{\text{MR}}^{(m)}\nonumber\\
&\qquad+\ldots\nonumber\\
&\qquad+\big(m(m-1)\,r_1r_2\big)^{N/2} \prod_i \theta_i \eta_i \cdot \Phi_{\text{HR}}^{(m)}.
\end{align}
Intriguingly, with two species of Grassmann coordinates, there appear Laughlin, Moore-Read, and Haldane-Rezayi states as expansion coefficients.
Each of them naturally appears in the following limits: the Laughlin state at $r_1,r_2\rightarrow 0$, the Moore-Read state at $r_1 \rightarrow \infty$ or $r_2 \rightarrow \infty$ with  $r_1 r_2$ fixed, and the Haldane-Rezayi state at $r_1,r_2 \rightarrow \infty$.
Now, let us move to the discussion of the VBS model with two species of Grassmann coordinates. The corresponding generalized AKLT state is
\begin{equation}
\Psi_{\rm AKLT}^{(M)}=\prod_{\langle ij\rangle}^z(u_iv_j-v_i u_j+r_1\theta_i\theta_j+r_2 \eta_i\eta_j)^M,
\end{equation}
and, for 1D spin chain, it is rewritten
\begin{align}
\Psi_{\text{AKLT}}^{(M)}&= \Phi_{\text{AKLT}}^{(M)}\cdot\exp\biggl(\!Mr_1\!\!\sum_{i}\!
\frac{\theta_i\theta_{i+1}}{u_iv_{i+1}-v_iu_{i+1}}\!\biggr)\label{extraAKLT2}\\
&\qquad\cdot\exp\biggl(\! Mr_2\!\!\sum_{i}\!\frac{\eta_i\eta_{i+1}}{u_iv_{i+1}-v_iu_{i+1}}
\!\biggr)\nonumber\\
&\qquad\cdot \exp\biggl(-M r_1r_2\sum_{i}\frac{ \theta_i\theta_{i+1}
\eta_i\eta_{i+1}}{(u_i v_{i+1}-v_i u_{i+1})^2}\biggr)\ .\nonumber
\end{align}
In the following, we concentrate on the case $M=2$.
The factor of the third exponent $\theta_i\theta_{i+1}\eta_i\eta_{i+1}/(u_i v_{i+1}-v_i u_{i+1})^2$  is  interpreted as the ``double-bond breaking operator'': it annihilates two valence bonds and creates two kinds of fermion pairs between $i$ and $i+1$ sites [Fig. \ref{bondbreaking2fig}].
Then, in Eq.(\ref{extraAKLT2}), there are two types of bond breaking operations, one of which is the single-bond breaking operations performed by first and second exponents, and the other is the double-bond breaking operation by the third exponent. With this interpretation, we have a nice graphical understanding of the expansion of the generalized AKLT state (see Fig. \ref{expSUSYAKLT2fig}).
As expected from the graphical representation, in the last terms of
 the order of $(r_1 r_2)^{L/2}$ there appear two fully dimerized states and two partially dimerized states. An explicit calculation yields
\begin{align}
& \int \prod_{i}d\theta_i d\eta_i \,\Psi_{\text{AKLT}}^{(M=2)}=\nonumber\\
&\qquad\quad  (2r_1 r_2)^{L/2}  \left(\prod_{i\atop \rm even}+\prod_{i\atop\rm odd}\right)(u_iv_{i+1}-v_iu_{i+1})^2\nonumber\\
&\qquad\quad-2^{L+1} (r_1 r_2)^{L/2} \prod_{i}(u_iv_{i+1}-v_i u_{i+1})\ ,
\label{lasttermdoublebond}
\end{align}
where once again we consider a ring of $L$ sites, with $L$ even.
Equation (\ref{lasttermdoublebond}) corresponds to the expression (\ref{appearanceofhr}) of the QHE. The first two terms on the RHS in (\ref{lasttermdoublebond}) denote the two fully dimerized states, while the last term on the RHS represents the two partially dimerized states. These fully and partially dimerized states are degenerate zero-energy eigenstates of the three-body truncated pseudopotential Hamiltonian, 
\begin{equation}
H_{D}=\sum_{i}\sum_{J=2}^{3} V_J P_J(i,i+1,i+2).
\end{equation}
The degeneracies may be resolved by adding terms involving other projection operators to the
Hamiltonian \cite{affleck1988vbg}.
Since the fully dimerized states in Eq.(\ref{lasttermdoublebond}) only take the spin magnitude $J=1$ for
groups of three consecutive sites, they are the zero-energy eigenstates of the Hamiltonian, 
\begin{align}
&H_{FD}=H_D+\sum_{i}V_{0}P_{0}(i,i+1,i+2)\nonumber\\
&~~~~~~=\sum_{i}\sum_{J\neq 1}V_JP_J(i,i+1,i+2)\ ,
\label{FDhamiltonian}
\end{align}
while the partially dimerized states are not.

Comparing the two expressions (\ref{lasttermdoublebond}) and (\ref{appearanceofhr}), one notices the
apparent analogies between the fully dimerized  double-bond states and the HR state.
As in the case of the dimerized single-bond state and the MR state, they share common features such as the
truncated pseudopotential Hamiltonians which render them exact ground states.  For the fully dimerized state,
the Hamiltonian is given by the three-body interaction form (\ref{FDhamiltonian}), while for the HR state,
it has a similar form ,
\begin{equation}
H_{\rm HR}=\sum_{i<j}\sum_{J=3(S-m)+3}^{3S} V_J P_J(i,j,k)\ ,
\end{equation}
with $S={1\over 2}\big(m(N-1)-2\big)$.

The generalization with more fermionic coordinates is a straightforward task. With $F$ species of fermionic coordinates, the SUSY AKLT state is generalized as
\begin{equation}
\Psi_{\rm AKLT}^{(M)}=\prod_{\langle ij\rangle }\Big(u_iv_j-v_i u_j+\sum_{f=1}^F r_f\theta^f_i\theta^f_j\Big)^M\!\!,
\end{equation}
and is rewritten as
\begin{align}
\Psi_{\rm AKLT}^{(M)}&=  \exp\biggl(\!M\sum_f^F r_f
\sum_{\langle ij\rangle}\frac{\theta^f_i\theta^f_j}{u_iv_j-v_iu_j}\biggr) \nonumber\\
&\quad \cdot\exp\biggl(\!-M\sum_{f<f'}^F r_f r_f'\sum_{\langle ij\rangle}\frac{\theta^f_i\theta^{f}_j\theta^{f'}_i\theta^{f'}_j}{(u_i v_j-v_i u_j)^2}\biggr)\nonumber\\
&\quad \cdot\exp\biggl(2M\!\!\!\!\sum_{f<f'<f''}^F \!\!\!\!r_f r_{f'}r_{f''}
\sum_{\langle ij\rangle}\frac{\theta^f_i\theta^{f}_j\theta^{f'}_i\!\theta^{f'}_j\!
\theta^{f''}_i\!\theta^{f''}_j  }{(u_i v_j-v_i u_j)^3}\biggr)\nonumber\\
&\quad\cdot\>\cdots\nonumber\\
&\quad\cdot \exp\biggl(\!(-1)^{F-1}(F-1)!\,M r_1r_2\cdots r_F\nonumber\\
&\qquad\qquad\cdot\sum_{\langle ij\rangle}\frac{\theta_i^1\theta_j^1\theta_i^2\theta_j^2\cdots \theta_i^F\theta_j^F}{(u_iv_j-v_iu_j)^F}\biggr)
\cdot\Phi_{\rm AKLT}^{(M)}.\label{exponetialmanyfermions}
\end{align}
As in the previous discussion, we consider the expansion of the exponentials in Eq.(\ref{exponetialmanyfermions}).
At the first term of the expansion, we obtain the original AKLT state with $S={1\over 2}zM$.  The last terms, of order $(M^F r_1 r_2\cdots r_F)^{L/2}$, represent a nearest neighbor RVB state with $S={1\over 2}(zM-F)$. For
the SVBS spin chain, the last terms are (fully and  partially) dimerized states that are degenerate zero-energy eigenstates of the three-body interaction Hamiltonian,
\begin{equation}
H_{D}=\sum_{i} \sum_{J=S+1}^{3S}V_J P_J(i,i+1,i+2)\ ,
\end{equation}
with $S={1\over 2}(2M-F)$. When $M=F$, the two degenerate fully dimerized states appear in the last terms, and are the zero-energy eigenstates of the truncated Hamiltonian
\begin{equation}
H_{FD}=\sum_{i} \sum_{J\neq S}^{3S}V_J P_J(i,i+1,i+2) \ ,
\end{equation}
with $S={1\over 2} M$.

\subsection{BCS aspects of the SVBS state}

In Sec.\ref{QHE}, we have mainly discussed the property of the SVBS
state in the two limits $r\rightarrow 0\,,\,\infty$ and found that the
$M=1$ SVBS spin chain produces the original AKLT state at
$r\rightarrow 0$, while the MG state at $r\rightarrow \infty$.
With finite $r$, the SVBS state contains a finite density of hole
pairs, and accordingly exhibits superconducting properties.
This state of affairs is familiar from the BCS state,
\begin{equation}
\big|{{\rm BCS}}\big\rangle=\prod_k \frac{1}{\sqrt{1+|g_k|^2}}\,
\big(1+g_k\, c^{\dagger}_{k}c^{\dagger}_{-k}\big)\,\big|{0}\big\rangle\ .
\end{equation}
As $g_k\rightarrow 0$, the BCS state is reduced to the vacuum,
while at $g_k\rightarrow\infty$, it becomes the completely
filled Fermi sphere.  For intermediate $g_k$, the $|{\rm BCS}\rangle$ describes
a state with off-diagonal long-ranged order. Then, one may conjecture the following
correspondences:
\begin{equation}
 g_k \leftrightarrow r,~~~~~~\big|{0}\big\rangle\leftrightarrow \Phi_{\rm AKLT}, ~~~~~\big|{F}\big\rangle\leftrightarrow \Phi_{\rm MG}\ .
\end{equation}
Interestingly, the BCS state exhibits a duality ($S$ duality, in terminology of high-energy theory) with respect
to the coherence factor,
\begin{equation}
g_k\leftrightarrow 1/g_k^*.
\label{sdualtrans}
\end{equation}
To see this, it is important to notice that the BCS state is represented in two ways,
\begin{align}
&\big|{{\rm BCS}}\big\rangle= \prod_k \frac{1}{\sqrt{1+|g_k|^2}}
\exp\big( g_k \,c_{k}^{\dagger}c^{\dagger}_{-k}\big){\big|{0}\big\rangle}
\nonumber\\
&~~~~~~~~~~=\prod_k \frac{1}{\sqrt{1+|g_k|^{-2}}}\,\exp\big( g_k^{-1}\, h_{k}^{\dagger}h^{\dagger}_{-k}\big)
{\big|{0}\big\rangle\!\!\big\rangle},
\label{bcsparticlepicture}
\end{align}
where $h_k$ represents the hole operator $h_k^{\dagger}=c_{-k}$, and $ \big|{0}\big\rangle\!\!\big\rangle$ is the hole vacuum, with $h_k\, \big|{0}\big\rangle\!\!\big\rangle =0$, namely, the fully occupied Fermi sphere
$ \big|{0}\big\rangle\!\!\big\rangle=\big|{{\rm F}}\big\rangle$. As seen in Eq.(\ref{bcsparticlepicture}), the two descriptions in terms of particle and hole operators are completely equivalent, and the duality physically
represents the particle-hole symmetry.  The order parameter
\begin{equation}
\Delta_k=\langle c^{\dagger}_{k}c^{\dagger}_{-k}\rangle=
\frac{g_k^*}{1+|g_k|^2}=\frac{1}{{g_k}+{g_k^*}^{-1}}\ ,
\label{BCSorderpara}
\end{equation}
manifestly reflects the dual structure of Eq.(\ref{sdualtrans}).  The order parameter thus vanishes in two limits:
the weak limit $g_k\rightarrow 0$, and the strong limit $g_k\rightarrow\infty$.  It takes its maximum value at
the self-dual point $|g_k|=1$.  The average occupancy of the momentum $k$ state, and its fluctuation, are
given by
\begin{subequations}
\begin{align}
&\langle n_k\rangle =\frac{|g_k|^2}{1+|g_k|^2},\\
&\langle (n_k-\langle{n_k}\rangle)^2\rangle = \frac{|g_k|^2}{(1+|g_k|^2)^2}=
\frac{1}{(g_k+{g_k^*}^{-1})(g_k^*+g_k^{-1})}.
\end{align}
\end{subequations}
The fluctuation, too, is maximalized at the self-dual point $|g_k|=1$.
As the duality is manifest in the BCS state and especially between
 $\big|{0}\big\rangle$ and $\big|{{\rm F}}\big\rangle$, one may speculate a hidden duality between the
 AKLT state and the MG state
\begin{equation}
\Phi_{\rm AKLT} \stackrel{{\rm dual?}}{\longleftrightarrow} \Phi_{\rm MG}.
\end{equation}
Indeed, the parameter-dependent terms in $\textsf{OSp}(1|2)$ Casimir operator, Eq.(\ref{susyspinspinint}),
are given by
\begin{equation}
-\frac{1}{4r}\,(a^{\dagger}_ib^{\dagger}_j-b^{\dagger}_ia^{\dagger}_j)\,f_if_j-\frac{r}{4}\,(a_ib_j-b_ia_j)\,
f_i^{\dagger}f^{\dagger}_j,
\end{equation}
which implies a duality
\begin{equation}
r\leftrightarrow {1}/{r} \quad,\quad
a_i b_j-b_i a_j \leftrightarrow f_i f_j\ .
\label{correpondenceabff}
\end{equation}
This is also the case {\it vis-a-vis\/} the truncated pseudopotential Hamiltonians for the SVBS
states. Physically, this duality corresponds to the interchange of VB and fermion pair,
in which case the SVBS state of Eq.(\ref{SAKLT}) is obviously invariant under the dual
transformation.  Though the VB and the fermion pair operators possess
same antisymmetric property with interchange of $i$ and $j$, their
squares exhibit different properties: the square of the VB is
non-zero, while the fermion pair vanishes.  More typically,
we cannot naively take the limit $r\rightarrow\infty$ in the SVBS
state, since in that limit, the SVBS state becomes
\begin{equation}
\Psi_{\rm AKLT}\rightarrow \prod_{\langle ij\rangle}\eta_i\,\eta_j=0,
\end{equation}
unlike the BCS state. Because of the asymmetric property between VB and fermion pair, the SVBS spin chain is not self-dual at the point $|r|=1$ and the order parameter [Eq.(\ref{swaveorderpara})] takes its maximum value
\begin{equation}
|\Delta_{\text{max}}|=(\sqrt{5}-2)\sqrt{\frac{2M(1+\sqrt{5})}{M+1}}
\end{equation}
at
\begin{equation}
|r|=\biggl(M+\frac{1}{2}\biggr)\sqrt{\frac{1+\sqrt{5}}{2M(M+1)}}.
\end{equation}
The expectation values for the boson number $n_{\rm b}(i)=a^{\dagger}_ia_i+b^{\dagger}_ib_i$ and the fermion number
$n_{\rm f}(i)=f^{\dagger}_if_i$
are calculated as
\begin{align}
&\langle n_{\rm b}\rangle =2M-1 +\frac{2M+1}{\sqrt{4M(M+1)(1+|r|^2)+1}}, \nonumber\\
&\langle n_{\rm f}\rangle = 1-\frac{2M+1}{\sqrt{4M(M+1)(1+|r|^2)+1}}.
\end{align}\label{expectnumber}
As expected, with increasing $|r|$, $\langle n_{\rm b}\rangle$ monotonically decreases, while $\langle n_{\rm f}\rangle $ monotonically increases.
The fluctuations for the boson number $\delta n_{\rm b}^2 =\langle{n^2_{\rm b}\rangle-\langle{n_{\rm b}}\rangle}^2$ and the fermion number $\delta n_{\rm f}=\langle n^2_{\rm f}\rangle-\langle n_{\rm f}\rangle^2$ are also evaluated as
\begin{align}
\delta n_{\rm b}^2=\delta n_{\rm f}^2&=x(1-x)\nonumber\\
x&=\frac{2M+1}{\sqrt{4M(M+1)(1+|r|^2)+1}}\ ,
\end{align}
and their maximum is $\delta n_{\rm b}=\delta n_{\rm f}={1\over 2}$ at $x={1\over 2}$, or
\begin{equation}
|r|=3\bigg(1+{1\over 4M(M+1)}\bigg)\ .
\end{equation}

\section{Hamiltonians for the SVBS state}\label{trancpseudohamil}

In Secs.\ref{CHAINS} and \ref{QHE}, we have studied the properties of the SVBS
state [Eq.(\ref{SAKLT})] and its relation to the Abelian and non-Abelian
fractional quantum Hall wave functions. To obtain a better
understanding of what physical systems the SVBS states describe, we shall in
this section construct a Hamiltonian for which the SVBS state is a unique ground state.

\subsection{Generic truncated pseudopotential Hamiltonian}\label{trancpseudohamilsvbs}
As mentioned in Sec.\ref{SUSY}, the SVBS state [Eq.(\ref{SAKLT})] 
is invariant under $\textsf{OSp}(1|2)$ transformations generated by
the parameter-dependent generators, $L_a$ and $K_{\mu}$, when
$x^2=-r$. Taking advantage of this symmetry, it is possible to construct
pseudopotential Hamiltonians for the SVBS states with
$\it{arbitrary}$ values of the parameter $r$. Truncated
pseudopotential Hamiltonians for the SVBS states [Eq.(\ref{SAKLT})] are
constructed by following the similar methods of the original AKLT model.
The superspin operator on site $i$,
$L_i=\frac{1}{2}(a^{\dagger}_i a_i+b^{\dagger}_i b_i+f^{\dagger}_i f_i)$, acts
the SVBS state to yield the eigenvalue $L=\frac{1}{2}zM.$ The
$z$ component of the bond superspin $J_{ij}^{z}=L^z(i)+L^z(j)=
\frac{1}{2}(a^{\dagger}_i a_i+a^{\dagger}_j a_j-b^{\dagger}_i b_i-b^{\dagger}_j b_j)$
counts the difference between the powers of $a$ and $b$ in the SVBS
state [Eq.(\ref{SAKLT})], and the maximal value of $J^z$ reads as
$J^{z}_{\text{max}}=(z-1)M=2L-M.$ Since the SVBS state is invariant
under  the $\textsf{OSp}(1|2)$ transformation, the maximal magnitude
of bond superspin is equal to that of its $z$-component, {\it i.e.}, 
$J_{\text{max}}=J^z_{\text{max}}.$ Thus, the SVBS state does not
contain any $\textsf{OSp}(1|2)$ angular-momentum components larger
than $J_{\text{max}}$ and is a zero-energy ground state of the
truncated pseudopotential Hamiltonian, 
\begin{equation}
H=\sum_{\langle ij\rangle}\sum_{J=J_{\text{max}}+{1\over 2}}^{2L}V_{J}\,{\mathbb P}_J(ij), \label{SUSYparentHamil}
\end{equation}
where $V_J$ are positive coefficients.
$\mathbb{P}_{J}(ij)$ is the projection operator made by $\textsf{OSp}(1|2)$ Casimir operators,
\begin{align}
&\mathbb{P}_J(ij)=\prod_{J'\neq J}^{2L}\frac{( K_A(i)+K_A(j) )^2-J'(J'+\frac{1}{2}) }{J(J+\frac{1}{2})-J'(J'+\frac{1}{2})} \nonumber\\
&~~~~~~~~=\prod_{J'\neq
J}^{2L}\frac{2K_A(i)K_A(j) +2L(L+\frac{1}{2})-J'(J'+\frac{1}{2})
}{J(J+\frac{1}{2})-J'(J'+\frac{1}{2})}, \label{projectiongene}
\end{align}
which projects to the two-site subspace of the bond superspin $J$.
Here, we have used $K_A^2(i)=K_A^2(j)=L(L+{1\over 2})$ with
$K_A^2=L_a^2+\epsilon_{\mu\nu}K_{\mu}K_{\nu}$. Apparently, the
projection operator [Eq.(\ref{projectiongene})] is $\textsf{OSp}(1|2)$
invariant, and hence the truncated pseudopotential Hamiltonian
(\ref{SUSYparentHamil}) as well. Following similar discussions in the AKLT
model, one may prove that the SVBS state is the unique zero-energy
eigenstate of the Hamiltonian (\ref{SUSYparentHamil}).

As an explicit example, it would be worthwhile to demonstrate the
truncated pseudopotential Hamiltonian for the $L=1$ SVBS spin chain.
With the $\textsf{OSp}(1|2)$ decomposition rule (\ref{osp1times1}), 
Eq.(\ref{SUSYparentHamil}) becomes
\begin{align}
&H_{\text{chain}}=\sum_i (V_{{3\over 2}} \,\mathbb{P}_{{3\over 2}}(i,i+1)+V_2 \,\mathbb{P}_2(i,i+1))\nonumber\\
&=\sum_i\biggl(\frac{32}{315}(V_{2}-{7}V_{{3\over 2}})(K_A(i)K_A(i+1) )^4\nonumber\\
&~~~~~~~~+\frac{16}{45}(V_{2}-{5}V_{{3\over 2}}) (K_A(i)K_A(i+1) )^3\nonumber\\
&~~~~~~~~
+\frac{2}{45}(9V_{2}-{7}V_{{3\over 2}}) (K_A(i)K_A(i+1) )^2\nonumber\\
&~~~~~~~~+\frac{1}{35}(5V_{2}+63V_{{3\over 2}}) K_A(i)K_A(i+1) +
V_{{3\over 2}}\biggr). \label{explicitSUSYAKLT}
\end{align}
In the special case $V_2=7V_{{3/ 2}}$, the first term on the last
RHS in Eq.(\ref{explicitSUSYAKLT}) vanishes, and
(\ref{explicitSUSYAKLT}) is reduced to
\begin{equation}
H_{\text{chain}} \rightarrow \frac{4}{45} \sum_i V_{{3\over 2}} \,\mathbb{P}_{{3/ 2}\oplus 2}(i,i+1),
\end{equation}
where $\mathbb{P}_{{3/ 2}\oplus 2}$ is the projection operator onto the space with
bond superspin ${3\over 2}$ or $2$,
\begin{align}
&\mathbb{P}_{{3\over 2}\oplus 2}(i,i+1)\nonumber\\
&= \prod_{J=0,{1}/{2},1}({(K_A(i)+K_A(i+1) )^2
  -J(J+\frac{1}{2})})\nonumber\\
&= ~~~8 (K_A(i)K_A(i+1) )^3 +
28(K_A(i)K_A(i+1) )^2\nonumber\\
&~~~~+\frac{63}{2}K_A(i)K_A(i+1) +\frac{45}{4}.
\end{align}

However, Hamiltonian (\ref{SUSYparentHamil}) cannot correspond to that of any physical system,
since it is non-Hermitian$^1$ {\footnotetext[1]{Though the Hamiltonian (\ref{SUSYparentHamil}) is non-Hermitian,
its eigenvalues are still real.
Recently, the study of such non-Hermitian Hamiltonians with real
eigenvalues has attracted much attentions \cite{bender2007msn}, and
the present Hamiltonian would be an interesting example.}} because of the term
$\epsilon_{\mu\nu}K_{\mu}K_{\nu}$, as mentioned in Sec.{\ref{SUSY}}.  To obtain a physical Hamiltonian for which the SVBS state
 is its unique ground state, one can replace the Hamiltonian
 (\ref{SUSYparentHamil}) by the following form:
\begin{equation}
H=\sum_{\langle ij\rangle}\sum_{J=J_{\text{max}}+{1\over 2}}^{2L}V_J\,\mathbb{P}_J^{\dagger}(ij)\,
\mathbb{P}_J(ij).
\label{hermitianHamil}
\end{equation}
in which $V_J>0$ just like in Eq.(\ref{SUSYparentHamil}). Here we would like
to make several comments on some properties of the Hermitian
Hamiltonian. First, the definition (\ref{hermitianHamil}) is a
natural generalization of the original pseudopotential Hamiltonian, since,
if the projection operators were Hermitian, with the property
$\mathbb{P}_J^{2}=\mathbb{P}_J$, Eq.(\ref{hermitianHamil}) would be reduced to the
original form (\ref{SUSYparentHamil}). Second, unlike the
non-Hermitian Hamiltonian (\ref{SUSYparentHamil}),
Eq.(\ref{hermitianHamil}) is {\it not} $\textsf{OSp}(1|2)$ SUSY
invariant, because the Hermitian conjugate of the
$\textsf{OSp}(1|2)$ Casimir operator contained in $\mathbb{P}^{\dagger}_J$ is no
longer invariant under the original $\textsf{OSp}(1|2)$
transformation. Consequently, the excitation spectrum of the
Hermitian Hamiltonian is not SUSY invariant, even though the ground state
remains is a SUSY singlet. Third, Hamiltonian
(\ref{hermitianHamil}) does not preserve the total fermion number
$N_{\rm f}=\sum_if_i^\dagger f_i$ since the Casimir operator
$(K_A(i)+K_A(j))^2$ contains pair-creation terms of fermions, as
shown in Appendix \ref{app:spinint}. This is in agreement with the
fermion number fluctuation in the SVBS state [Eq.(\ref{SAKLT})].
Physically, such a pseudo-potential Hamiltonian describes some
interacting electron system coupled with a superconducting bath,
which provide a particle bath through proximity effect.

Since $\mathbb{P}_J^{\dagger}(ij)\,\mathbb{P}_J(ij)$ is always non-negative, it is
straightforward to prove that $H\big|{G}\big\rangle=0$ for a state $\big|{G}\big\rangle$ if
and only if $\mathbb{P}_J(ij)=0$ for all sites and all $J_{\rm max}<J\leq
2L$. Consequently, if the SVBS state is the only zero-energy
eigenstate of Hamiltonian (\ref{SUSYparentHamil}), it must also be
the unique ground state of the Hermitian Hamiltonian
(\ref{hermitianHamil}).  One can then prove the
SVBS state to be the unique ground state of Hamiltonian
(\ref{hermitianHamil}) following exactly the same procedure as
AKLT's original work \cite{affleck1987rrv,affleck1988vbg}. We will leave the detail of this
proof as the task of Appendix \ref{proofHamil}. Here, we sketch the proof for $L=1$ superspin chain.
Let $\big|{\Psi_G}\big\rangle$ be a ground state of Hamiltonian (\ref{hermitianHamil}) and satisfy the equation
\begin{equation}
H\big|{\Psi_G}\big\rangle=0.
\end{equation}
Then,
\begin{subequations}
\begin{align}
&\big\langle{\Psi_G}\big|H\big|{\Psi_G}\big\rangle=0 \nonumber\\
&\Rightarrow  \big\langle{\Psi_G}\big|\mathbb{P}^{\dagger}_J(ij)\,\mathbb{P}_J(ij)\big|{\Psi_G}\big\rangle_{J>J_{\text{max}}}=0\label{secondcons}\\
&\Rightarrow \mathbb{P}_J(ij)\big|{\Psi_G}\big\rangle_{J>J_{\text{max}}}=0,
\end{align}
\end{subequations}
where in the second arrow [Eq.(\ref{secondcons})] we have used that $V_J$ in Eq.(\ref{SUSYparentHamil}) satisfy $V_J>0$.
Meanwhile, if $\big|{\Psi_G}\big\rangle$ is annihilated by the projection operator, {\it i.e.\/},  if
\begin{equation}
\mathbb{P}_J(ij)\big|{\Psi_G}\big\rangle_{J>J_{\text{max}}}=0,
\label{conditionaboutuniq}
\end{equation}
then it immediately follows that $H\big|{\Psi_G}\big\rangle=0$.
Thus, the condition (\ref{conditionaboutuniq}) is the necessary and sufficient condition such that the $\big|\Psi_G\big\rangle$ is the ground state of the Hamiltonian (\ref{hermitianHamil}).
We use the condition (\ref{conditionaboutuniq}) to show $\big|{\Psi_G}\big\rangle$ is the unique ground state of the Hamiltonian.
For $L=1$, the condition (\ref{conditionaboutuniq}) is given by
\begin{equation}
\mathbb{P}_{{3\over 2}}(i,i+1)\big|{\Psi_G}\big\rangle=\mathbb{P}_{2}(i,i+1)\big|{\Psi_G}\big\rangle=0.
\label{conditionsforuniqueness}
\end{equation}
As we assumed, there is superspin 1 on each site of the chain, and therefore, if the two superspins on sites $i$ and $i+1$ did not combine a $\textsf{OSp}(1|2)$ singlet, their bond superspin inevitably would exceed $J_{\text{max}}=1$ due to the $\textsf{OSp}(1|2)$ decomposition rule [Eq.(\ref{osp1times1})]. This observation holds for bond superspins on arbitrary two neighboring sites.
Then, on any two neighboring sites, the bond superspin should form a
$\textsf{OSp}(1|2)$ singlet, and the ``bulk''  ground state is given by the products of neighboring $\textsf{OSp}(1|2)$ singlet states.
Hence, with periodic boundary, it is apparent that the SVBS chain state [Eq.(\ref{generalakltchainfunc})] is the unique ground state.
With open boundaries, there are ninefold quasi-degenerate ground states
 corresponding to directions of the superspins on two ends, 
\begin{equation}
\big|{\Psi_G}\big\rangle_{\mu\nu}=\psi_{\mu,0}^{\dagger}\cdot \prod_{i=1}^{L-1}(a^{\dagger}_ib^{\dagger}_{i+1}-b_i^{\dagger}a_{i+1}^{\dagger}+rf^{\dagger}_if^{\dagger}_{i+1})\cdot \psi_{\nu,L}^{\dagger}\big|{0}\big\rangle,
\end{equation}
where ${\mu,\nu}=a,b,f$.
These ninefold quasi-degenerate states generally take different expectation values for local observable $A$,
\begin{equation}
\big\langle{A}\big\rangle_{\mu\nu}
=\frac{\big\langle{\Psi_G}\big|A\big|{\Psi_G}\big\rangle_{\mu\nu}}{\big\langle{\Psi_G}\big|{\Psi_G}\big\rangle_{\mu\nu}}.
\label{expectlocalobs}
\end{equation}
However, as in the original AKLT case \cite{affleck1988vbg}, the different energy eigenvalues converge in the infinite chain limit as we shall see below.
Suppose the length of the chain $N$ (from site 0 to site $N$), and  $A$ takes its support in $\{l,\ldots,N-l\}$  $(l\ll N)$.
First, we discuss the integration of the numerator of  
Eq.(\ref{expectlocalobs}) from one end (site $0$) to site $l$.
The inner products of the superspin states at site $0$ are denoted as
\begin{equation}
\alpha_0+\beta_0\,\hat{n}^z_0+\gamma_0\,\theta_0\,\theta^*_0.
\end{equation}
The self-inner products of $u_0$, $v_0$, and $\theta_0$ correspond to $(\alpha_0,\beta_0,\gamma_0)=
({1\over 2},{1\over 2},0),~({1\over 2},-{1\over 2},0)$, 
and $(0,0,1)$, respectively.  The integration from site $j$ to site $j+1$ induces the transformation:
\begin{equation}
\begin{pmatrix}
\alpha_j\\
\beta_j\\
\gamma_j
\end{pmatrix}
\rightarrow
\begin{pmatrix}
\alpha_{j+1}\\
\beta_{j+1}\\
\gamma_{j+1}
\end{pmatrix}
= \begin{pmatrix}
\frac{3}{2} & 0 & \frac{1}{2}\\
0 & -\frac{1}{2} & 0\\
|r|^2 & 0 & 0
\end{pmatrix}
\begin{pmatrix}
\alpha_j\\
\beta_j\\
\gamma_j
\end{pmatrix}.
\end{equation}
The three eigenvalues of the transfer matrix are given by $\lambda_{\pm}=({3\pm\sqrt{9+8|r|^2}})/{4}$ and $-{1\over 2}$, and then, at $l\rightarrow \infty$, the product of the transfer matrices provides
\begin{equation}
T^l\rightarrow \frac{\lambda^l_+}{\lambda_+ - \lambda_-}
\begin{pmatrix}
\lambda_+ & 0 & \lambda_-\\
0 & 0 & 0 \\
-\lambda_+ & 0 & -\lambda_-
\end{pmatrix}.
\end{equation}
Then, if there is $u_0$ or $v_0$ at site $0$, we have a factor $(1-\theta_l\theta_l^*)\,{\lambda_+^{l+1}}/{2(\lambda_+-\lambda_-)}$ at site $l$,
while if  $\theta_0$, we have  a different value $(1-\theta_l\theta_l^*)\,{\lambda_+^{l}\lambda_-}/(\lambda_+-\lambda_-)$, but
the results only differ by the scaling factor, and such difference is not relevant to $\big\langle A\big\rangle_{\mu\nu}$ since the scaling factor is canceled between the numerator and the denominator in Eq.(\ref{expectlocalobs}).
Thus, the integration is not relevant to directions of the superspin at site $0$ in  the infinite limit.
The integration from the other end (site $N$) to site $N-l$ gives same consequence.
Then, regardless of directions of superspins on boundaries, the expectation value of any local observable provides a unique value
\begin{equation}
\big\langle A\big\rangle_{\mu\nu}\rightarrow \big\langle A\big\rangle,
\end{equation}
and, in this sense, the ninefold quasi-degenerate SUSY ground states converge to the unique ground state on infinite chain.

\subsection{Another Hamiltonian for fixed total fermion number}\label{MODEL}



In this subsection, we will show an alternative Hamiltonian for
the simplest $L=1$ case, which is not constructed from the ${\rm
OSp(1|2)}$ Casimir operators but has the advantage of respecting
fermion number conservation. Motivated by the three-site
Hamiltonian known for Majumdar-Ghosh spin chain \cite{majumdar1969nnn}, here
we construct a Hamiltonian with both two-site and three-site
terms, for which the projection of the SVBS state [Eq.(\ref{SAKLT})] to
a fixed total fermion number is a unique ground state. Such AKLT
states with fixed fermion number have appeared in each order of
the expansion of the SVBS state as seen in Sec.\ref{physicalintSVBS}.
 For simplicity,  we will focus on the $M=1$ case, {\em i.e.}, a chain with $S=1$ or
$S={1\over 2}$ on each site.
We will first write down the form of the Hamiltonian before
analyzing the physical meaning of each term.
\begin{eqnarray}
H&=&H_t+H_V+H_U-\mu \sum_if_i^\dagger f_i\ ,\nonumber\\
H_V&=&\sum_i\left(V_{{3\over 2}}\,P_{{3\over 2}}(i,i+1)+V_{2}\,P_{2}(i,i+1)\right)\ ,\nonumber\\
H_U&=&\sum_iU_{{3\over 2}}\,P_{{3\over 2}}(i,i+1,i+2)\ ,\nonumber\\
H_t&=&-t\sum_i\left(\Delta_{i,i+1}\,\Delta^\dagger_{i+1,i+2}+h.c.\right)\label{H3site}
\end{eqnarray}
with $P_J(i,i+1)$ and $P_J(i,i+1,i+2)$ the two-site and three-site
projections to total $\textsf{SU}(2)$ spin $J$ states,
respectively, and $\Delta_{i,i+1}=f_i^\dagger
f_{i+1}^\dagger\left(a_ib_{i+1}-b_ia_{i+1}\right)$ the
annihilation operator of a Cooper pair. It should be noticed that
the Hamiltonian is defined in the Hilbert space satisfying the
constraint $a_i^\dagger a_i+b_i^\dagger b_i+f_i^\dagger
f_i=2,~\forall i$. The coefficients $V_2,~U_{3\over 2},~t$ are all
positive. The chemical-potential term $-\mu\sum_if_i^\dagger f_i$
determines the fermion number in the ground states.

\begin{figure}[!t]
\centering
\includegraphics[width=8.7cm]{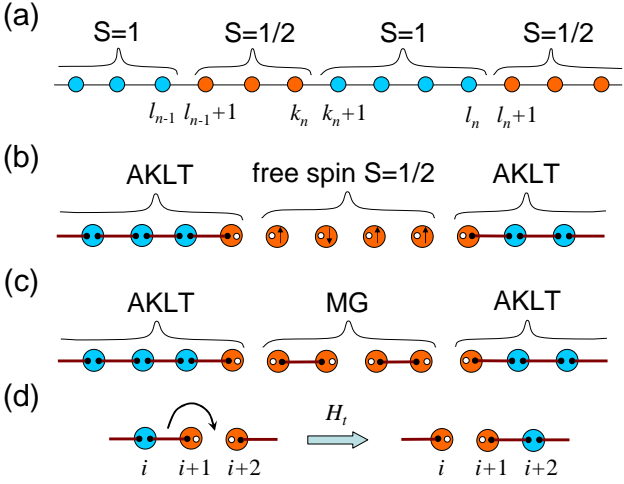}
\caption {\label{fig:H} (a) Schematic picture of a spin
configuration of the SVBS chain. The blue or orange sites are spin-$1$
and spin-${1\over 2}$, respectively. $k_n$ and $l_n$ label the last site of
each $S={1\over 2}$ ($S=1$) segment. (b) Schematic picture of the ground
states of $H_V$ in Eq.(\ref{H3site}). Each solid line stands for a
nearest-neighbor singlet pair. The spin of $S={1\over 2}$ sites are free
except for the neighbor sites of the $S=1$ segments. (c) Schematic
picture of the ground states of $H_V+H_U$. The $S=1$ sites form AKLT
state (terminated by a $S={1\over 2}$ site) and the $S={1\over 2}$ sites form
dimerized MG state. For a fixed configuration of $S=1$ and $S={1\over 2}$
sites, the ground state is unique. (d) The effect of the hopping
term $H_t$, which hops a nearest-neighbor singlet from $i,i+1$ link
to $i+1,i+2$ link, or vice versa.}
\end{figure}

To understand the ground-state property of Hamiltonian
(\ref{H3site}), we start from the interaction terms $H_V+H_U$. Since
$H_V+H_U$ preserves the fermion number $n^h_i=f_i^\dagger f_i$ on
each lattice site, one can focus on studying its matrix element
within a subspace defined by fixed eigenvalue of $n^h_i$. For any
given configuration $\left\{n_i^h\right\}_{i=1}^{N}$, the 1D
chain can be viewed as consecutive staggered sectors of spin-$1$ and
spin-${1\over 2}$ chains, as shown in Fig. \ref{fig:H}(a). When
$\left\{n_i^h\right\}$ satisfies
\begin{eqnarray}
n_i^h=\begin{cases}0 &{\rm for}\ k_n<i\leq l_n\\
1&{\rm for}\ l_n<i\leq k_{n+1}\ ,\end{cases} \nonumber
\end{eqnarray}
with $n\in\{1,\ldots,M\}$,
the chain consists of $M$ spin-$1$ chains with lengths $l_n-k_n$ and
$M$ spin-${1\over 2}$ chains with lengths $k_{n+1}-l_n$. (Here
$k_{M+1}=k_1$.)

Now we consider the effect of $H_V$ and $H_U$ on such a spin chain.
Firstly, the two-site projector $P_2(i,i+1)$ is nontrivial only when
there are no fermion on the two sites $(i,\,i+1)$, because the total
spin is automatically smaller than $2$ if there are one or two holes
on these two sites. Therefore, the $V_2$ term in $H_V$ is an AKLT
Hamiltonian acting on the disconnected spin-$1$ segments $k_n<i\leq
l_n$. Thus we immediately know that the $V_2$ term takes the minimal
eigenvalue of zero if the spin-$1$ segments $k_n<i\leq l_n$ are all
spin-$1$ AKLT spin chains.

Second, the two-site projector $P_{{3/2}}(i,i+1)$ is nontrivial only
when there is one fermion on the two sites $(i,i+1$), {\em i.e.},
$n_i^h+n_{i+1}^h=1$. For these sites, the requirement
$P_{{3/ 2}}(i,i+1)=0$ leads to singlet pair between the free $S={1\over 2}$
spin at the end of the AKLT spin-$1$ chain and the neighbor
spin-${1\over 2}$ site. This requirement automatically fixes the length
of each spin-${1\over 2}$ segment $k_{n+1}-l_n$ to be $\geq 2$. Other
spin-${1\over 2}$ sites which are not neighbor of spin-$1$ site are not
affected by $H_V$. In summary, the spin configuration with vanishing
eigenvalue of $H_V$ is shown in Fig. \ref{fig:H} (b).

Third, the three-site projector $P_{{3/ 2}}(i,i+1,i+2)$ is nontrivial
only when there are one or three fermions on the three sites
$(i,\,i+1,\,i+2)$. When there are one fermion on the three sites, it can
be proved that any spin configuration which satisfy $H_V=0$ also
satisfy $H_U=0$ automatically. Thus we only need to consider the
effect of $H_U$ on the sites with three fermions, {\em i.e.}, three
consecutive sites with $n_i^h=n_{i+1}^h=n_{i+2}^h=1$. In other
words, $H_U$ is exactly the Majumdar-Ghosh Hamiltonian for the
$S={1\over 2}$ segments. As known from the work of Majumdar and Ghosh, the
ground-state requirement $H_U=0$ can only be satisfied by the two
valence bond solid states, with spin singlet pairs between each two
nearest-neighbor sites. Moreover, the connect condition to the $S=1$
segments will pick one of the two VBS states, as shown in Fig.
\ref{fig:H} (c). (Also, the length of each $S={1\over 2}$ segment is
automatically required to be even, in order to form singlet pairs.)

In summary, the ground state of interaction terms $H_U+H_V$ is
uniquely determined for a given distribution of $S=1$ and $S={1\over 2}$
sites. Now we consider the effect of the hopping term $H_t$. The
operator $\Delta_{i,i+1}$ annihilates a singlet pair and creates two
fermions on $i$ and $i+1$ sites. Thus
$\Delta_{i,i+1}\Delta^\dagger_{i+1,i+2}$ flips a singlet from
$i,i+1$ link to $i+1,i+2$ link. Notice that
$\Delta^\dagger_{i+1,i+2}=\left(a_{i+1}^\dagger
b_{i+2}^\dagger-b_{i+1}^\dagger
a_{i+2}^\dagger\right)f_{i+2}f_{i+1}$, we know that the term
$\Delta_{i,i+1}\Delta^\dagger_{i+1,i+2}$ has nonzero matrix element
only if $n_{i+1}^h=n_{i+2}^h=1,~n_i^h=0$. In other words, $H_t$ only
acts on the interface sites between $S=1$ and $S={1\over 2}$ segments.
Moreover, in the ground-state manifold of $H_V+H_U$, the effect of
$H_t$ is simply hopping of a nearest-neighbor singlet, as shown in
Fig. \ref{fig:H} (d). From this picture we know that $H_t$ preserves
a ground state of $H_V+H_U$ in the ground-state manifold.
Consequently, $H_t$ lifts the degeneracy of the ground-state
manifold of $H_V+H_U$. The lowest energy state determined by $H_t$
for a fixed total fermion number is obviously the equal weight
superposition of all the spin configurations satisfying $H_V+H_U=0$,
which is exactly the SVBS state [Eq.(\ref{SAKLT})] projected to a fixed
fermion number,
\begin{eqnarray}
\left|G_N\right\rangle=P_N\prod_{i} \big(a_i^\dagger
b_{i+1}^\dagger-b_i^\dagger a_{i+1}^\dagger + f^{\dagger}_i
f^{\dagger}_j\big)\big|{0}\big\rangle\label{ProjectSVBS}
\end{eqnarray}
It should be noticed that $|G_N\rangle$ is nonvanishing only when $N$
is even, otherwise the ground state cannot be a spin singlet. As the
last step, the fermion number $N$ for which the state $|G_N\rangle$
has lowest energy can be tuned by the chemical-potential term
$-\mu\sum_if_i^\dagger f_i$. It is possible that for some $\mu$ the
ground state contains odd number of fermions, which thus cannot be
SVBS state.

In conclusion, we have shown that Hamiltonian (\ref{H3site}) has
the SVBS state [Eq.(\ref{ProjectSVBS})] as its unique ground state, as
long as $t,V_{{3/2}},V_2,U_{{3/ 2}}>0$ and the chemical potential is
chosen properly so that the ground state has even number of
fermions. We have also confirmed this fact numerically by
diagonalizing Hamiltonian (\ref{H3site}) for up to five sites with
periodic boundary condition and calculating the overlap between the
numerical ground-state wave function and the projected SVBS state
[Eq.(\ref{ProjectSVBS})]. Within numerical accuracy, the ground state of
Hamiltonian (\ref{H3site}) for even total fermion number is unique
and always given by the SVBS state [Eq.(\ref{ProjectSVBS})].



\section{Conclusions}
In conclusion we have constructed the supersymmetric
generalization of the valence bond solid states. In one dimension,
these SVBS states smoothly interpolates between the integer and
half-integer VBS states, and they represent superconducting
valence bond liquid states. We also constructed microscopic
Hamiltonians for which these states are the exact quantum ground
states. We show that the SVBS states are analogous to bosonic
Pfaffian states of the quantum Hall effect, in precisely the same
sense as the analogy between the VBS states and the Laughlin
quantum Hall states. Our work also provides a precise mathematical
realization of some ideas in strongly correlated systems, in the
sense that the doped valence bond liquid states are naturally
superconducting, and that the superconducting states can be
obtained from a symmetry rotation, in our case a supersymmetric
rotation, of the quantum antiferromagnetic ground states. For the
future, we propose to focus on the two- and higher-dimensional
versions of the SVBS states. Given the analogies between the SVBS
states and the Pfaffian states in the quantum Hall effect, it
would also be interesting to explore the possibility of
non-Abelian statistics of the elementary excitations.

\section*{Acknowledgements}

D.P.A. would like to thank Andrei Bernevig for useful discussions. K.H. 
acknowledges Mitsuhiro Arikawa, Keisuke Totsuka, and Masatoshi Sato
for helpful comments and telling references. D.P.A. and K.H. are glad to
thank the condensed-matter group of Stanford Institute for
Theoretical Physics, where this work is initiated, for warm
hospitality, a stimulating atmosphere, and supports for sabbatical
and travels. This work is supported by the NSF under Grant No. 
DMR-0342832 and the U.S. Department of Energy, Office of Basic Energy
Sciences under Contact No. DE-AC03-76SF00515. The work of K.H. was also
partially supported by Sumitomo foundation.

\appendix

\section{$\textsf{OSp}(1|2)$ and $\textsf{SU}(2|1)$  Algebras}\label{superalgebras}

 Here, we review basic properties of $\textsf{OSp}(1|2)$ and $\textsf{SU}(2|1)$ algebras with emphasis on their relation to Schwinger boson and slave fermion formalism. The $\textsf{OSp}(1|2)$ algebra consists of five generators, $L_A=L_a (a=1,2,3)$ and $L_{\mu} (\mu=\theta_1,\theta_2)$ that satisfy
\begin{align}
&[L_a,L_b]=i\epsilon_{abd}L_c,\nonumber\\
&[L_a,L_{\mu}]=\frac{1}{2}(\sigma_a)_{\nu\mu}L_{\nu},
\nonumber\\
&\{L_{\mu},L_{\nu}\}=\frac{1}{2}(\epsilon\sigma_a)_{\mu\nu}L_a,
\label{osp12algebra}
\end{align}
where  $\sigma_a$ are Pauli matrices, and $\epsilon$ is the $2\times 2$ antisymmetric matrix $\epsilon=i\sigma_2$.
Equation (\ref{osp12algebra}) suggests that $L_a$ transform as $\textsf{SU}(2)$ vector and $L_{\mu}$ $\textsf{SU}(2)$ spinor. The Casimir operator for the $\textsf{OSp}(1|2)$ group is given by
\begin{equation}
C=L_A L_A\equiv L_aL_a+\epsilon_{\mu\nu}L_{\mu}L_{\nu},
\label{casimirforsop12}
\end{equation}
and its eigenvalue is $L(L+{1\over 2})$ with integer of half-integer $L$.
$L$ is referred to as superspin and characterizes the irreducible
representations of $\textsf{OSp}(1|2)$. The dimension of irreducible representation with
superspin $L$ is $4L+1$, $2L+1$ of which is the $\textsf{SU}(2)$ spin $L$
representation, and the remaining $2L$ is $\textsf{SU}(2)$ spin $L-{1\over 2}$. Specifically, the $\textsf{OSp}(1|2)$ fundamental representation $L={1\over 2}$ is three-component spinor and the corresponding $\textsf{OSp}(1|2)$ generators are the following $3\times 3$ matrices:
\begin{equation}
l_a=\frac{1}{2}
\begin{pmatrix}
\sigma_a & 0 \\
0 & 0
\end{pmatrix},~~~~
l_{\mu}= \frac{1}{2}
\begin{pmatrix}
0 & \tau_{\mu} \\
-(\epsilon\tau_{\mu})^t & 0
\end{pmatrix},\label{fundosp12}
\end{equation}
with $\tau_1=(1,0)^t$ and $\tau_2=(0,1)^t$. The irreducible
decomposition for superspin representations is given by
\begin{equation}
L\otimes L' = |L-L'|\oplus |L-L'|+{1}/{2} \oplus |L-L'|+1 \oplus \cdots
\oplus L+L'. \label{irreducibledecoosp}
\end{equation}
Unlike the $\textsf{SU}(2)$ decomposition rule,
the superspins on the RHS differ by ${1\over 2}$.

The $\textsf{SU}(2|1)$ or $\textsf{OSp}(2|2)$ algebra consists of eight generators; $L_a,L_{\mu}$
[$\textsf{OSp}(1|2)$ generators], $D_{\mu}$ and $\Gamma$ that satisfy
\begin{align}
&[L_a,D_{\mu}]=\frac{1}{2}(\sigma_a)_{\nu\mu}D_{\nu},\nonumber\\
&\{D_{\mu},D_{\nu}\}=-\frac{1}{2}(\epsilon\sigma_a)_{\mu\nu}L_a,\nonumber\\
&\{L_{\mu},D_{\nu}\}=-\frac{1}{4}\epsilon_{\mu\nu}\Gamma,\nonumber\\
&[L_a,\Gamma]=0,\nonumber\\
&[L_{\mu},\Gamma]=-D_{\mu},\nonumber\\
&[D_{\mu},\Gamma]=-L_{\mu}. \label{su21algebra}
\end{align}
With Eq.(\ref{fundosp12}), the simplest matrix realization for Eq.(\ref{su21algebra}) is given by
\begin{equation}
d_{\mu}= \frac{1}{2}
\begin{pmatrix}
0 & -\tau_{\mu} \\
-(\epsilon\tau_{\mu})^t & 0
\end{pmatrix},
~~~\gamma=
\begin{pmatrix}
1 & 0 & 0 \\
0 & 1 & 0 \\
0 & 0 & 2
\end{pmatrix}.
\end{equation}
As the Schwinger
particle used in the $\textsf{SU}(2)$ spin formalism, slave fermion is introduced in  the superspin formalism.
We denote Schwinger bosons as $\textsf{SU}(2)$ spinor $b_{\mu}=(a,b)$, and slave fermion as $\textsf{SU}(2)$ singlet $f$, and
they satisfy the commutation relations:
$[a,a^{\dagger}]=[b,b^{\dagger}]=\{f,f^{\dagger}\}=1$. The $\textsf{SU}(2|1)$ operators are represented as
\begin{align}
&L_a=\psi^{\dagger} l_a \psi=\frac{1}{2}(\sigma_a)_{\mu\nu}b_{\mu}^{\dagger}b_{\nu},\nonumber\\
&L_{\mu}=\psi^{\dagger} l_{\mu} {\psi}=\frac{1}{2}(b_{\mu}^{\dagger}f+\epsilon_{\mu\nu}f^{\dagger}b_{\nu}),\nonumber\\
&D_{\mu}=\psi^{\dagger} d_{\mu} {\psi}=\frac{1}{2}(-b_{\mu}^{\dagger}f+\epsilon_{\mu\nu}f^{\dagger}b_{\nu}),\nonumber\\
&\Gamma=\psi^{\dagger}\gamma
\psi=a^{\dagger}a+b^{\dagger}b+2f^{\dagger}f,
\end{align}
where $\psi=(a,b,f)^t=(b_1,b_2,f)^t$.
$L_{\mu}$ and $D_{\mu}$ are not Hermitian in the conventional
sense, while with the definition of the superstar conjugation
$\ddagger$
\begin{equation}
(f^{\ddagger})^{\ddagger}=-f,~~(f_1f_2)^{\ddagger}=f_1^{\ddagger}f_2^{\ddagger},\label{defpseudoconj}
\end{equation}
they become pseudo-Hermitian operators
\begin{equation}
L_{\mu}^{\ddagger}=\epsilon_{\mu\nu}L_{\nu},~~D_{\mu}^{\ddagger}=-\epsilon_{\mu\nu}D_{\nu}.
\end{equation}
(The detail definition of the superstar conjugation can be referred to Ref. 
\cite{Dictionarysuperalgebras}.) In the slave fermion representation, the $\textsf{OSp}(1|2)$ Casimir operator (\ref{casimirforsop12}) is rephrased as
\begin{equation}
C=\frac{a^{\dagger}a+b^{\dagger}b+f^{\dagger}f}{2}\biggl(\frac{a^{\dagger}a+b^{\dagger}b+f^{\dagger}f}{2}+\frac{1}{2}\biggr),
\end{equation}
and the superspin magnitude corresponds to the half of the total particle number
\begin{equation}
L=\frac{1}{2}(a^{\dagger}a+b^{\dagger}b+f^{\dagger}f).
\end{equation}

We introduce a complex parameter $x$ to define one-parameter family
of fermionic generators made by $L_{\mu}$ and $D_{\mu}$
\begin{align}
&K_{\mu}=\frac{1}{2x}(L_{\mu}-D_{\mu})+\frac{x}{2}(L_{\mu}+D_{\mu})
\nonumber\\
&~~~~=\frac{1}{2}
\begin{pmatrix}
a\\
b\\
f
\end{pmatrix}^{\dagger}
\begin{pmatrix}
0 & \frac{1}{x}\tau_{\mu}\\
-x (\epsilon\tau_{\mu})^t & 0
\end{pmatrix}
\begin{pmatrix}
a\\
b\\
f
\end{pmatrix}\nonumber\\
&~~~~=\frac{1}{2x}b_{\mu}^{\dagger}f+\frac{x}{2}\epsilon_{\mu\nu}f^{\dagger}b_{\nu}.
\end{align}
At $x=1$, $K_{\mu}$ is reduced to $L_{\mu}$, and at $x=i$,
$K_{\mu}=iD_{\mu}$.
Though $K_{\mu}$ depends on the parameter $x$, interestingly, $L_a$ and $K_{\mu}$ satisfy the parameter $\it{independent}$ $\textsf{OSp}(1|2)$ algebraic relations
\begin{align}
&[L_a,L_b]=i\epsilon_{abc}L_c,\nonumber\\
&[L_a,K_{\mu}]=\frac{1}{2}(\sigma_a)_{\nu\mu}K_{\nu},\nonumber\\
&\{K_{\mu},K_{\nu}\}=\frac{1}{2}(\epsilon\sigma_a)_{\mu\nu}L_a.
\end{align}\label{newosp12alge}
The Casimir operator is given by
\begin{align}
&K_A^2\equiv L_a^2+\epsilon_{\mu\nu}K_{\mu}K_{\nu}\nonumber\\
&=
L_a^2+\biggl(\frac{x}{2}+\frac{1}{2x}\biggr)^2\epsilon_{\mu\nu}L_{\mu}L_{\nu}
+\biggl(\frac{x}{2}-\frac{1}{2x}\biggr)^2\epsilon_{\mu\nu}D_{\mu}D_{\nu},\label{newosp12casimir00}
\end{align}
which, in the slave fermion representation, is expressed as
\begin{equation}
K_A^2=
\frac{a^{\dagger}a+b^{\dagger}b+f^{\dagger}f}{2}\biggl(\frac{a^{\dagger}a+b^{\dagger}b+f^{\dagger}f}{2}+\frac{1}{2}\biggl).
\label{newosp12casimir}
\end{equation}
Again,  the parameter $x$  does not appear in Eq.(\ref{newosp12casimir}) and the eigenvalues of the Casimir operator are given by $L(L+{1}/{2})$ for any of the one-parameter family.

\section{SUSY Spin-Spin Interactions}\label{app:spinint}

Here, we discuss several properties of the $\textsf{OSp}(1|2)$ spin-spin interaction
\begin{equation}
 K_A(i)K_A(j) =L_a(i)L_a(j)+\epsilon_{\mu\nu}K_{\mu}(i)K_{\nu}(j),
 \label{defomedspin-spin}
\end{equation}
where $i$ and $j$ represent the sites on which superspins are
defined. Since the SUSY spin-spin interaction operator commutes
with the superspin-magnitude operator $L_i={1}/{2}(a^{\dagger}a+b^{\dagger}b+f^{\dagger}f)_i$,
the SUSY spin-spin interaction does not change the magnitude of the superspin on each site.
The bosonic spin-spin interaction part of Eq.\ref{defomedspin-spin})
gives the $\textsf{SU}(2)$ spin-spin interaction
\begin{align}
&L_a(i)L_a(j)\nonumber\\
&=\frac{1}{2}a^{\dagger}_i a_j  b^{\dagger}_j b_i  +
 \frac{1}{2}a_j^{\dagger}a_i b_i^{\dagger}b_j +\frac{1}{4}(a^{\dagger}a-b^{\dagger}b)_i(a^{\dagger}a-b^{\dagger}b)_j,\label{spinspininteori}
\end{align}
while the fermionic spin-spin interaction part of  
Eq.(\ref{defomedspin-spin}) provides
\begin{align}
&\epsilon_{\mu\nu}K_{\mu}(i)K_{\nu}(j)\nonumber\\
&=\biggl(\frac{x}{2}+\frac{1}{2x}\biggr)^2\epsilon_{\mu\nu}L_{\mu}(i)L_{\nu}(j) +\biggl(\frac{x}{2}-\frac{1}{2x}\biggr)^2\epsilon_{\mu\nu}D_{\mu}(i)D_{\nu}(j) \nonumber\\
&+\biggl(\frac{x}{2}+\frac{1}{2x}\biggr)\biggl(\frac{x}{2}-\frac{1}{2x}\biggr)\epsilon_{\mu\nu}(L_{\mu}(i)D_{\nu}(j)+D_{\mu}(i)L_{\nu}(j)),
\label{newosp12casimir0}
\end{align}
and, in the Slave fermion representation, expressed as
\begin{align}
&\epsilon_{\mu\nu}K_{\mu}(i)K_{\nu}(j)\nonumber\\
&=\frac{1}{4x^2}(a^{\dagger}_ib^{\dagger}_j-b^{\dagger}_ia^{\dagger}_j)f_if_j+\frac{x^2}{4}(a_ib_j-b_ia_j)f_i^{\dagger}f^{\dagger}_j\nonumber\\
&+\frac{1}{4}(a_i^{\dagger}a_j+b_i^{\dagger}b_j) f_j^{\dagger}f_i
+\frac{1}{4}(a^{\dagger}_j a_i+b^{\dagger}_jb_i)f^{\dagger}_i f_j.
\label{defOSpshwingwe}
\end{align}
The first two terms on the RHS in Eq.(\ref{defOSpshwingwe}) are particular interactions existing in the $\textsf{OSp}(1|2)$ spin-spin term.
They violate the total fermion number conservation, and represents hole-pair annihilating and creating interactions.
Besides, they are not Hermitian even at $x=1$. [However, at $x=1$, they  are pseudo-Hermitian with the definition of the superstar conjugation (\ref{defpseudoconj}).]
 The last two terms represent interchange of fermion and  boson between $i$ and $j$ sites. Meanwhile, the $\textsf{SU}(2|1)$ spin-spin interaction is given by
\begin{align}
&L_a(i)L_a(j)+\epsilon_{\mu\nu}L_{\mu}(i)L_{\nu}(j)-\epsilon_{\mu\nu}D_{\mu}(i)D_{\nu}(j)-\frac{1}{4}\Gamma(i)\Gamma(j)\nonumber\\
&=\frac{1}{2}a^{\dagger}_i a_j  b^{\dagger}_j b_i  +
 \frac{1}{2}a_j^{\dagger}a_i b_i^{\dagger}b_j +\frac{1}{4}(a^{\dagger}a-b^{\dagger}b)_i(a^{\dagger}a-b^{\dagger}b)_j\nonumber\\
&+ \frac{1}{2} (a_j^{\dagger}a_i+b^{\dagger}_j b_i)f_i^{\dagger}f_j+
  \frac{1}{2} (a_i^{\dagger}a_j+b^{\dagger}_i b_j)f_j^{\dagger}f_i   \nonumber\\
&- \frac{1}{4}(a^{\dagger}a+b^{\dagger}b+2f^{\dagger}f)_i
(a^{\dagger}a+b^{\dagger}b+2f^{\dagger}f)_j, \label{su21intspin1}
\end{align}
and is the component of the SUSY $t-J$ model Hamiltonian.
It should be noted that the particular hole-pair creating and annihilating terms in 
Eq.(\ref{defOSpshwingwe}) do not exist in the $\textsf{SU}(2|1)$ spin-spin interaction.

Though the $\textsf{OSp}(1|2)$-invariant spin-spin interaction is not Hermitian,
its eigenvalues are real and do not depend on the parameter $x$.
Indeed, with two-body operator $K_A(i,j)=K_A(i)+K_A(j)$, the $\textsf{OSp}(1|2)$ spin-spin interaction (\ref{defomedspin-spin}) is simply rewritten as
\begin{equation}
K_A(i)K_A(j) =\frac{1}{2}K_A(i,j)^2-\frac{1}{2}K_A(i)^2-\frac{1}{2}K_A(j)^2\ ,
\label{spinspinrewrite}
\end{equation}
and its eigenvalues are
\begin{equation}
E=\frac{1}{2}J(J+\frac{1}{2}) - \frac{1}{2}L_i(L_i+\frac{1}{2}) -
\frac{1}{2}L_j(L_j+\frac{1}{2})\ , \label{eigenstateenergyofdefosp}
\end{equation}
where $J$, $L_i$, and $L_j$ are the Casimir indexes for $K_A(i,j)$, $K_A(i)$, and $K_A(j)$, respectively.

One may confirm above features with a low energy example. The two-body states $\big|{J, J_3}\big\rangle$ made by $L_i={1\over 2}$ and  $L_j={1\over 2}$, carry the $\textsf{OSp}(1|2)$ Casimir indexes $J=0{1\over 2},1$ by the decomposition rule (\ref{irreducibledecoosp}).
The $J=0$ sector consists of
\begin{equation}
 \big|{0,0}\big\rangle=(a_i^{\dagger}b_j^{\dagger}-b_i^{\dagger}a_j^{\dagger}-x^2 f_i^{\dagger}f_j^{\dagger})\big|{0}\big\rangle\ . \label{deformed00}
\end{equation}
This is the $\textsf{OSp}(1|2)$ singlet state, and is the ``component'' of the
SVBS state [Eq.(\ref{SAKLT})]. The $J={1\over 2}$ sector consists of
\begin{align}
&\big|{\frac{1}{2},\frac{1}{2}}\big\rangle=(a^{\dagger}_if_j^{\dagger}-f_i^{\dagger}a_j^{\dagger})\big|{0}\big\rangle,\nonumber\\
&\big|{\frac{1}{2},0}\big\rangle=(a^{\dagger}_ib_j^{\dagger}-b_i^{\dagger}a_j^{\dagger}-2x^2f_i^{\dagger}f_j^{\dagger})\big|{0}\big\rangle,\label{deformed1/20}\nonumber\\
&\big|{\frac{1}{2},-{1\over 2}}\big\rangle
=(b^{\dagger}_if_j^{\dagger}-f_i^{\dagger}b_j^{\dagger})\big|{0}\big\rangle\ .
\end{align}
Similarly, the $J=1$ sector consists of
\begin{align}
 &\big|{1,1}\big\rangle=a_i^{\dagger}a_j^{\dagger}\big|{0}\big\rangle,\nonumber\\
 &\big|{1,{1\over 2}}\big\rangle=( a_i^{\dagger}f_j^{\dagger}+f_i^{\dagger}a_j^{\dagger})\big|{0}\big\rangle,\nonumber\\
 &\big|{1,0}\big\rangle= (a_i^{\dagger}b_j^{\dagger}+b_i^{\dagger}a_j^{\dagger})\big|{0}\big\rangle,\nonumber\\
 &\big|{1,-{1\over 2}}\big\rangle=( b_i^{\dagger}f_j^{\dagger}+f_i^{\dagger}b_j^{\dagger})\big|{0}\big\rangle,\nonumber\\
 &\big|{1,-1}\big\rangle=b_i^{\dagger}b_j^{\dagger}\big|{0}\big\rangle.\label{deformed1/2}
 \end{align}
Equation (\ref{eigenstateenergyofdefosp}) suggests that  $J=0$, $J={1\over 2}$, 
and $J=1$ sectors carry eigenvalues $E=-{1\over 2}, -{1\over 4}$, and ${1\over 4}$,
respectively. By applying the $\textsf{OSp}(1|2)$-invariant spin-spin interaction operator
to these states, one may confirm such parameter-independent eigenvalues are obtained.
The parameter dependence appears only in the eigenstates $\big|{0,0}\big\rangle$ and
$\big|{{1\over 2},0}\big\rangle$, as found in Eqs.(\ref{deformed00}) and (\ref{deformed1/20}).

\section{Proof of the SVBS state as unique ground state of
Hamiltonian (\ref{hermitianHamil})}\label{proofHamil}

In this appendix we will prove that the SVBS state [Eq.(\ref{SAKLT})] is
unique ground state of Hamiltonian (\ref{hermitianHamil}). The
procedure of this proof is a straightforward supersymmetric
generalization of AKLT's original work \cite{affleck1987rrv,affleck1988vbg}. To finish the
proof, we need to consider the open-boundary condition. The boson
and fermion can be written in a $\textsf{OSp}(1|2)$ spinor
$\psi_i=(a_i,b_i,f_i)$, and the SVBS state can be written as
\begin{eqnarray} \big|{{\rm SVBS}}\big\rangle=\prod_i\left(\psi_{i\mu}^\dagger
C^{\mu\nu}\psi_{i+1,\nu}^\dagger\right)^M\big|{0}\big\rangle, \end{eqnarray}
where
\begin{equation}
C^{\mu\nu}=
\begin{pmatrix}
0 & 1 & 0 \\
-1 & 0 & 0\\
0 & 0 & r
\end{pmatrix}.
\end{equation}

For a open-boundary chain with length $L$, the definition needs to
be modified by \begin{eqnarray}
\big|{{\rm SVBS};\left\{\mu_s,\nu_t\right\}}\big\rangle&=&\left(\prod_{s=1}^M\psi^\dagger_{1\mu_s}\right)
\prod_{i=1}^{L-1}\left(\psi_{i\sigma}^\dagger
C^{\sigma\tau}\psi_{i+1,\tau}^\dagger\right)^M\nonumber\\
& & \cdot\left(\prod_{t=1}^M\psi^\dagger_{L\nu_t}\right)\big|{0}\big\rangle \equiv
\hat{\Omega}_{\mu_s\nu_t}\big|{0}\big\rangle\label{SVBSobc} \end{eqnarray} in which
$\hat{\Omega}_{\mu_s\nu_t}\equiv\hat{\Omega}_{\mu_1\mu_2\cdots\mu_M;
\nu_1\nu_2\cdots\nu_M}$
is symmetric under the permutations $(\mu_1\mu_2\cdots\mu_M)$ and
$(\nu_1\nu_2\cdots\nu_M)$. In other words, the state
$\big|{{\rm SVBS};\left\{\mu_s,\nu_t\right\}}\big\rangle$ carries the ${\textsf{OSp}(1|2)}$
representation $\frac{M}2\otimes \frac{M}2$. For the open-boundary
system, we have the following lemma:

\begin{itemize}
\item {\bf Lemma 1}. On an open-boundary chain with length $L$, if a state $\big|{\Psi}\big\rangle$ satisfies
$P_{i,i+1}^N\big|{\Psi}\big\rangle=0,~\forall\ i=1,2,...,L-1,~N=M+{1\over 2},\ldots,2M$,
then the state is a superposition of the SVBS states
(\ref{SVBSobc}), {\em i.e.},
\begin{eqnarray}
\big|{\Psi}\big\rangle=A^{\mu_s\nu_t}\big|{{\rm SVBS},\left\{\mu_s,\nu_t\right\}}\big\rangle,~\exists
A^{\mu_s\nu_t} \end{eqnarray}
\end{itemize}

Lemma 1 can be proved by induction as follows: 

(1) In the two-site case $L=2$, the states in the Hilbert space are
classified by the superspin as \begin{eqnarray} M\otimes
M=0\oplus\frac12\oplus1\oplus...\oplus 2M. \end{eqnarray} The requirement
$\mathbb{P}_{12}^N\big|{\Psi}\big\rangle=0,~\forall\ N=M+{1\over 2},\ldots,2M$ requires the state to
stay in the sub-Hilbert space of
$0\oplus\frac12\oplus1\oplus\cdots\oplus M$ which has a dimension of
$1+3+\ldots+(4M+1)=(2M+1)^2$. On the other hand, the $(2M+1)^2$ states
$\big|{{\rm SVBS},\left\{\mu_s,\nu_t\right\}}\big\rangle$ are linearly independent and satisfy the constraint.  Consequently, the states
$\big|{{\rm SVBS},\left\{\mu_s,\nu_t\right\}}\big\rangle$ span a complete basis of the ground state Hilbert space. In other words, the lemma 1 for $L=2$ is proved.

(2) An arbitrary state $\big|{\Psi}\big\rangle_{1,L+1}$ in the Hilbert space of a
length $L+1$ chain can always be expanded as
$\big|{\Psi}\big\rangle_{1,L+1}=\sum_{n,m}\big|{n}\big\rangle_{1,L}\Psi_{nm}\otimes\big|{m}\big\rangle_{L+1}$,
with $\big|{n}\big\rangle_{1,L}$ and $\big|{m}\big\rangle_{L+1}$ an arbitrary set of basis
states for the Hilbert subspace of the first $L$ sites and that of the
last site.  By an SVD decomposition of the matrix $\Psi_{nm}$, one
can always obtain the form \begin{eqnarray}
\big|{\Psi}\big\rangle_{1,L+1}=\sum_k\lambda_k\big|{W_k}\big\rangle_{1,L}\otimes\big|{S_k}\big\rangle_{L+1}\ ,
\end{eqnarray} where $\big|W_k\big\rangle_{1,L}$ are orthogonal states in the Hilbert
space of a length-$L$ chain, and $\big|{S_k}\big\rangle$ are orthogonal states in
the Hilbert space of the $L+1$th site. The coefficients
$\lambda_k>0$. If $\mathbb{P}_{i,i+1}^N\big|\Psi\big\rangle_{1,L+1}=0$ for
$i=1,2,\ldots,L-1$, we have
\begin{align}
0&={\rm norm}\left[{\mathbb{P}_{i,i+1}^N\sum_k\lambda_k\big|W_k\big\rangle_{1,L}\otimes\big|{S_k}\big\rangle_{L+1}}\right]\nonumber\\
\Rightarrow\quad 0&=\sum_k\lambda_k^2\big\langle{W_k}\big|\mathbb{P}_{i,i+1}^{N^\dagger}
\mathbb{P}_{i,i+1}^N\big|W_k\big\rangle_{1,L}\nonumber\\
\Rightarrow\quad 0&=\mathbb{P}_{i,i+1}^N\big|W_k\big\rangle_{1,L}\nonumber\\
\Rightarrow\quad 0&=\big|W_k\big\rangle_{1,L}=A_k^{\mu_s\nu_t}\big|{{\rm SVBS},\left\{\mu_s,\nu_t\right\}}\big\rangle_{1,L}
\end{align}
The last step is inductive, assuming the result holds true for a system of $L$ sites.
Thus the state $\big|{\Psi}\big\rangle_{1,L+1}$ is written as
\begin{eqnarray}
\big|\Psi\big\rangle_{1,L+1}=\sum_k\lambda_kA_k^{\mu_s\nu_t}B_k^{\sigma_k\tau_l}\hat{\Omega}_{\mu_s\nu_t}^{1,L}
\hat{\Omega}_{\sigma_k\tau_l}^{L+1}\big|{0}\big\rangle \end{eqnarray} in which
$\hat{\Omega}^{L+1}_{\sigma_k\tau_l}=\prod_{k=1}^M\psi^\dagger_{L+1,\sigma_k}\prod_{l=1}^M\psi^\dagger_{L+1,\tau_l}
\big|{0}\big\rangle$. 
The indices $(\nu_s,\sigma_k)$ carry the representation $\frac{M}2\otimes
\frac{M}2$, which can be decomposed into irreducible representations
as $\frac{M}2\otimes \frac{M}2=0\oplus \frac12\oplus\cdots\oplus M$.
Such a decomposition can be expressed as \begin{eqnarray}
\big|\Psi\big\rangle_{1,L+1}=\sum_{N=0}^M\sum_{n=-N}^{N}
F^{\mu_s\tau_l}_{Nn}C_{Nn}^{\nu_t\sigma_k}\hat{\Omega}_{\mu_s\nu_t}^{1,L}
\hat{\Omega}_{\sigma_k\tau_l}^{L+1}\big|{0}\big\rangle\label{PsiLplus1} \end{eqnarray} in
which $C_{Nn}^{\nu_t\sigma_k},~n=-N,-N+{1\over 2},...,N$ are the
$3j$-symbols carrying the representation of $\frac{\bar{M}}2\otimes
\frac{\bar{M}}2\otimes N$. Thus in the state
$C_{Nn}^{\nu_t\sigma_k}\hat{\Omega}_{\mu_s\nu_t}^{1,L}
\hat{\Omega}_{\sigma_k\tau_l}^{L+1}\big|{0}\big\rangle$, the sites $L$ and $L+1$
carry the representation $\frac{M}2\otimes N\otimes \frac{M}2$.
Thus we know that the maximal total ${\textsf{OSp}(1|2)}$ ``angular
momentum" of these two sites is $M+N$. Consequently, the requirement
$\mathbb{P}_{L,L+1}^N\big|\Psi\big\rangle_{1,L+1}=0,~N>M$ requires that only $N=n=0$
terms are nonzero in Eq.(\ref{PsiLplus1}). In other words, the
state can be written as \begin{eqnarray}
\big|\Psi\big\rangle_{1,L+1}=F^{\mu_s\tau_l}_{00}C_{00}^{\nu_t\sigma_k}
\hat{\Omega}_{\mu_s\nu_t}^{1,L}
\hat{\Omega}_{\sigma_k\tau_l}^{L+1}\big|{0}\big\rangle\ . \end{eqnarray}
Moreover, the
coefficient $C_{00}^{\nu_t\sigma_k}$ can be expressed as
\begin{eqnarray}
C_{00}^{\nu_t\sigma_k}=\mathcal{S}\left[{\prod_{s=1}^MC^{\nu_s\sigma_s}}\right]
\end{eqnarray}
in which $\mathcal{S}\left[{\cdots}\right]$ stands for symmetrization over
the indices $\left\{\nu_t\right\}$ and $\left\{\sigma_k\right\}$, respectively. By the
definition of $\Omega_{\mu_s\nu_t}^{1,L}$ in Eq.(\ref{SVBSobc}), it
is straightforward to show that $C_{00}^{\nu_t\sigma_k}
\hat{\Omega}_{\mu_s\nu_t}^{1,L}
\hat{\Omega}_{\sigma_k\tau_l}^{L+1}=\Omega^{1,L+1}_{\mu_s\tau_l}$,
so that \begin{eqnarray}
\big|\Psi\big\rangle_{1,L+1}=F^{\mu_s\tau_l}_{00}\hat{\Omega}^{1,L+1}_{\mu_s\tau_l}\big|{0}\big\rangle=
F^{\mu_s\tau_l}_{00}\big|{{\rm SVBS},\left\{\mu_s,\tau_l\right\}}\big\rangle \ .\nonumber\\
\end{eqnarray}

In summary, we have proved lemma 1 by induction. By making use of
lemma 1, it is straightforward to prove that the SVBS state
[Eq.(\ref{SAKLT})] to be the unique ground state of Hamiltonian
(\ref{hermitianHamil}). First of all, it is easy to see that for any
physical state $\big|\Psi\big\rangle$, $\big\langle{\Psi}\big|H\big|\Psi\big\rangle=
\sum_{i}\sum_{N=M+{1\over 2}}^{2M}V_N\cdot
{\rm norm}\left(\mathbb{P}_{i,i+1}^N\big|\Psi\big\rangle\right)\geq 0$.
Since the SVBS state
[Eq.(\ref{SAKLT})] satisfies $H\big|{{\rm SVBS}}\big\rangle=0$, we know that it is a ground
state of Hamiltonian (\ref{hermitianHamil}). On the other hand,
if there is another state $\big|{G}\big\rangle$ satisfying $H\big|G\big\rangle=0$, we have
\begin{eqnarray} {\rm
norm}\left(\mathbb{P}_{i,i+1}^N\big|G\big\rangle\right)&=&0\nonumber\\
\Rightarrow\quad\mathbb{P}_{i,i+1}^N\big|G\big\rangle&=&0,~\forall i,~\forall M<N\leq 2M. \end{eqnarray}
Consider a chain with $L$ sites and periodic boundary condition.
According to lemma 1, the conditions $\mathbb{P}_{i,i+1}^N\big|G\big\rangle=0$ for
$i=1,2,\ldots,L-1$ lead to \begin{eqnarray}
\big|G\big\rangle=A^{\mu_s\nu_t}\hat{\Omega}^{1,L}_{\mu_s\nu_t}\big|{0}\big\rangle.\nonumber\end{eqnarray}
In the same way as has been used in the proof of lemma 1, the
coefficient $A^{\mu_s\nu_t}$ can be decomposed into different
irreducible representations as
\begin{eqnarray}
A^{\mu_s\nu_t}=\sum_{N=0}^M\sum_{n=-N}^NF_{Nn}C_{Nn}^{\mu_s\nu_t}\ .
\end{eqnarray}
Applying the condition $\mathbb{P}_{L,1}^N\big|G\big\rangle=0$ to the state
$\big|G\big\rangle=\sum_{N,n}F_{Nn}C_{Nn}^{\mu_s\nu_t}
\hat{\Omega}^{1,L}_{\mu_s\nu_t}\big|{0}\big\rangle$
we obtain $F_{Nn}=0$ for all $N\neq 0$. Thus we have proved that
\begin{eqnarray}
\big|G\big\rangle=C_{00}^{\mu_s\nu_t}\hat{\Omega}^{1,L}_{\mu_s\nu_t}\big|{0}\big\rangle
=\big|{\rm SVBS}\big\rangle\ .
\end{eqnarray}

In summary, the state $\big|{\rm SVBS}\big\rangle$ in Eq.(\ref{SAKLT}) is the
unique ground state of the generalized pseudo-potential Hamiltonian
(\ref{hermitianHamil}).

\vfill\eject

\bibliography{References}

\end{document}